\documentclass[12pt]{article}

\usepackage{amsmath}
\usepackage{amssymb}
\usepackage{a4wide}
\usepackage{dsfont}
\usepackage{color}

\begin{document}

\newcommand{\club}{$\clubsuit \ $}
\newcommand{\mclub}{$ \clubsuit \clubsuit \ $}

\newcommand{\bul}{$\P \ $}
\newcommand{\mbul}{$\P \P \ $}
\newcommand{\mmclub}{{\;  \clubsuit \;}}
\newcommand{\mmbul}{\P \ }

\newcommand{\pic}{$\ \spadesuit$}
\newcommand{\mpic}{\ \spadesuit}
\newcommand{\mmpic}{$\ \spadesuit\spadesuit$}
\newcommand{\para}{\P\ }
\newcommand{\mpara}{\mbox{\P}\ }
\newcommand{\mmpara}{\P\P\ }

\newcommand{\nc}{\newcommand}
\nc{\Id}{{\mathchoice {\rm 1\mskip-4mu l} {\rm 1\mskip-4mu l}{\rm 1\mskip-4.5mu l} {\rm 1\mskip-5mu l}}}

\newcommand{\hx}{\hat X}
\newcommand{\hp}{\hat P}
\newcommand{\ha}{\hat a}

\newcommand{\Bc}{ {\cal B} }

\newcommand{\asto}{\star_{\Omega}}

\newcommand{\br}{\bf R}
\newcommand{\bc}{\bf C}
\newcommand{\hs}{\cal H}

\newcommand{\ds}{\displaystyle}
\newcommand{\ri}{{\rm i}}
\newcommand{\re}{{\rm e}}
\newtheorem{theo}{Theorem}[section]
\newtheorem{lem}{Lemma}[section]
\newtheorem{prop}{Proposition}[section]
\newtheorem{cor}{Corollary}[section]
\newtheorem{defin}{Definition}[section]


\hfill LYCEN 2007-19
\vskip 0.07truecm

\thispagestyle{empty}

\bigskip

\begin{center}
{\bf \Huge{Magnetic fields in }}
\end{center}
\begin{center}
{\bf \Huge{noncommutative quantum mechanics}}
\end{center}

\bigskip

\centerline{Dedicated to the memory of 
{\bf Julius Wess}\footnote{We 
dedicate these notes to the memory of Julius Wess
whose scientific work was largely devoted to the study 
of gauge fields and whose latest interests concerned 
physical theories on noncommutative space. 
The fundamental and inspiring contributions of Julius to Theoretical Physics 
will always bear with us, but his great kindness, his clear 
and enthusiastic 
presentations, and his precious advice will be missed 
by all those who had the chance to meet him.} }

\bigskip
\bigskip

\centerline{\Large {Fran\c cois Delduc}{$^{\ast}$},
{Quentin Duret}{$^{\dagger}$},}

\medskip

\centerline{\Large {Fran\c cois Gieres}$\,$\footnote{Corresponding 
author: {\tt gieres@ipnl.in2p3.fr}}{$^{, \, \ddagger}$},  
{Matthieu Lefran\c cois}{$^{\ddagger}$}}

\bigskip
\bigskip

\centerline{
{$^{\ast}$}
{\it Universit\'e de Lyon, Laboratoire de Physique, UMR 5672} }
\centerline{{\it
\'Ecole Normale Sup\'erieure de Lyon,
46 all\'ee d'Italie,}}
\centerline{{\it F-69364 Lyon cedex 07 (France)}}

\bigskip

\centerline{
{$^{\dagger}$}
{\it Laboratoire de Physique Th\'eorique et Hautes \'Energies, UMR 7589} }
\centerline{{\it (CNRS / Universit\'e P. et M. Curie
Paris 6 / Universit\'eD. Diderot Paris 7),}}
\centerline{{\it  Tour 24-25, 5eme \'etage, BP 126, 4 place Jussieu,}}
\centerline{{\it F-75252 Paris Cedex 05 (France)}}

\bigskip

\centerline{$^{\ddagger}$ {\it Universit\'e de Lyon,
Institut de Physique Nucl\'eaire de Lyon,}}
\centerline{{\it Universit\'e Lyon 1 and CNRS/IN2P3, Bat. P. Dirac,}}
\centerline{{\it 4 rue Enrico Fermi, F-69622 Villeurbanne Cedex (France)}}

\bigskip
\bigskip
\bigskip

\begin{abstract}
We discuss various descriptions of a quantum
particle on noncommutative space in a (possibly non-constant)
magnetic field.
We have tried to present the basic facts
in a unified and synthetic manner,
and to clarify the relationship between various
approaches and results that are scattered in the literature. 

\end{abstract}

\newpage

\thispagestyle{empty}
\tableofcontents

\newpage
\setcounter{page}{1}

\section{Introduction and overview}

\noindent
{\bf What noncommutativity is about:}
Let us assume that
the components $\hx_1, \dots , \hx_d$ of the
quantum mechanical position operator in $d$-dimensional space
do not commute with each other, but rather satisfy the
commutation relations
$[ \hx _i , \hx _j ] = \ri \theta_{ij} \Id$ where
$\theta_{ij} = - \theta_{ji}$ are real numbers
(at least one of which is non-zero).
The Cauchy-Schwarz inequality then implies
the uncertainty relations
$(\Delta _{\psi}  \hx _i )( \Delta _{\psi} \hx _j )
\geq {1 \over 2} |\theta_{ij}|$,
i.e. the particle described by the wave function $\psi$
cannot be localized in a precise way.
This uncertainty relation for the position implies
a certain fuzziness of points in space:
one says that the space is {\em fuzzy, pointless} or that it has a {\em lattice,
quantum} or {\em noncommutative structure}~\cite{madore}.
Obviously, 
the noncommutativity can only manifest itself
if the configuration
space is at least of dimension two. Thus, it represents a
deformation of the classical theory 
which is quite different from
the deformation  
by quantization which concerns all dimensions
and which amounts to introducing a cellular structure 
(parametrized by $\hbar$) in phase space.

\medskip
\noindent
{\bf History of the subject:}
The idea of a fuzzy configuration space 
looks quite interesting for
quantum field theories since it may help to avoid, or
at least ameliorate, the problem of short-distance singularities, i.e.
ultraviolet divergences.
Indeed\footnote{The history of the subject has been traced back
by J.~Wess and is reported upon in references~\cite{jackiw, calmet}.},
this argument has already been put forward in 1930 by
Heisenberg~\cite{heisenberg} and the message was successively
passed on to Peierls, Pauli~\cite{pauli}, Oppenheimer and Snyder
(who was a student of the latter).
In 1933, Peierls worked out a quantum mechanical application
concerning a particle in a constant magnetic field
(see appendix)~\cite{peierls}
and, in 1947, Snyder considered
noncommuting coordinates on space-time in order to discard the
ultraviolet divergences in quantum field theory without destroying the
Lorentz covariance~\cite{Snyder:1946qz}.
In the sequel, the latter work was not further exploited
due to the remarkable success of the renormalization scheme
in quantum electrodynamics.
Recently, noncommuting coordinates have regained interest,
in particular in the framework of superstring theories
and of quantum gravity.
Roughly speaking, the intrinsic
length scale of the strings induces a noncommutative structure on
space-time at low scales.
(For a review of these topics, and in particular of 
quantum field theory on noncommutative space-time
and its renormalization, see for instance 
references~\cite{nekra}-\cite{vincent}.) 
Since quantum mechanics may be regarded as a field theory 
in zero spatial dimensions or as a non-relativistic  
description of the one-particle sector
of field theory, it is well suited as a toy model
for the introduction of noncommuting coordinates.

\medskip
\noindent
{\bf Definition of NCQM:}
Quantum mechanics on noncommutative configuration
space is generally referred to as
noncommutative quantum mechanics ({\bf NCQM}).
Somewhat more generally, one can consider the situation
where the commutators for
coordinates and momenta are non-canonical.
After some precursory work dealing with noncommutative
phase space variables in quantum mechanics~\cite{dunne}-\cite{duval},
NCQM has been defined in a simple and direct manner
using different approaches~\cite{mezin}-\cite{nair},
and some basic physical models have been
studied~\cite{mezin}-\cite{horvathy}.
In the sequel,
various aspects of the subject have been further
elaborated upon and the literature has grown to a
considerable size over the last years\footnote{We apologize
to the authors whose work is not explicitly 
cited here and we refer to~\cite{inprepa}
for a more complete list of references.}.  
The study of exactly solvable models of NCQM
should hopefully lead to a better understanding
of some issues in noncommutative field theory.
There are some
potential physical applications which include
condensed matter physics (quantum Hall effect, superfluidity,...)
and spinning particles.
By considering classical Poisson brackets
(instead of commutators)
and Hamilton's equations
of motion, one can also discuss {\em noncommutative classical mechanics},
as well as applications thereof like 
magneto-hydrodynamics~\cite{guralnik, jackiw}.

\medskip
\noindent
{\bf About the physical interpretation and applications:}
Since a cellular structure in configuration space
is not observed at macroscopic scales, 
the noncommutativity parameters $\theta_{ij}$
should only manifest themselves at
a length scale which is quite small
compared to some basic length scale like the Planck length
$\sqrt{\hbar G / c^3}$~\cite{madore}.
(In the context of field theory, one usually writes
$\theta _{ij} = {1 \over \Lambda _{NC} ^2}
\tilde{\theta} _{ij}$,
where the $\tilde{\theta} _{ij}$
are dimensionless and of order $1$,
so that $\Lambda _{NC}$ represents a characteristic energy scale
for the noncommutative theory which is necessarily quite large.)
Thus, noncommutativity of space may be related to gravity
at very short distances and NCQM may be 
regarded as a deformation
of classical mechanics that is independent
of the deformation by quantization.
From this point of view, a fully fledged theory
of quantum gravity  should provide a fuller understanding
of the noncommutativity of space.

In the framework of noncommutative classical or quantum mechanics, 
the parameters $\theta_{ij}$ admit many close analogies
with a constant magnetic field
both from the algebraic and dynamical
points of view~\cite{smail}.
Such an effective magnetic field, which is necessarily quite small,
may play a role as primordial magnetic field
in cosmological dynamics~\cite{aca3}.

Though various {\em physical applications} of NCQM 
have been advocated
(in particular for quantum mechanical systems
coupled to a constant magnetic field),
the only experimental signature of
noncommuting spatial coordinates 
which is currently available 
appears to be the approximate noncommutativity
appearing in the Landau problem
for the limiting case of a
very strong magnetic field -- see appendix.
(This is a realization of noncommutativity which
is analogous to the one in string theory).

\medskip
\noindent
{\bf Scope of the present paper:}
The essential features of NCQM can be exhibited
by considering
the following three instances to which we limit ourselves
in the present paper.
\begin{enumerate}
\item
We focus on {\bf flat $2$-dimensional space}
so that the antisymmetric matrix
$(\theta _{ij})$ is given by
$\theta _{ij} = \theta \varepsilon_{ij}$
where $\theta$ is a real number and  $\varepsilon_{ij}$
are the components of the antisymmetric tensor
normalized by $\varepsilon_{12} = 1$.
\item
We assume that the fundamental {\bf commutator algebra}
has the following form
(which is not the most general form that one can consider for 
NCQM~\cite{hatzi}):
\begin{equation}
\label{nccr-plane}
[ {\hx}_1 , {\hx}_2 ] = \ri \theta \Id
\, , \quad
[ {\hp}_1 , {\hp}_2 ] = \ri B \Id
\, , \quad
[ {\hx}_i , {\hp}_j ] = \ri \delta_{ij} \Id
\, .
\end{equation}
Here, the real constant $B$
(which measures the noncommutativity of momenta)
describes a {\em constant}
magnetic field that is perpendicular to the $\hx_1 \hx_2$-plane.
In equations (\ref{nccr-plane}) and in the following,
the variables $\hx_i, \hp_j$
(which are to be viewed as self-adjoint Hilbert space operators)
are denoted by a hat so as to distinguish them from the basic
operators $X_i, P_j$ of standard quantum mechanics
which satisfy the Heisenberg algebra,
i.e. the canonical commutation relations (CCR's)
\begin{equation}
\label{heisen}
[ {X}_1 , {X}_2 ]  = 0 = [ {P}_1 , {P}_2 ]
\, , \qquad
[ {X}_i , {P}_j ] = \ri \hbar \delta_{ij} \Id
\, .
\end{equation}
As we will see, one can express the operators $\hx_1, \hx_2,
\hp_1,  \hp_2$ as linear combinations of $X_1, X_2, P_1, P_2$ with
coefficients which depend on the {\em noncommutativity parameters}
$\theta$ and $B$ (that describe the deviation from the canonical
commutators). We note that the parameters $\theta$ and $B$ in
equations (\ref{nccr-plane}) are not necessarily independent of
each other and we will see that their interplay yields some
particularly interesting results~\cite{nair, bellucci, duvalhor,
smail}.
\item
We are interested in the case of a {\bf possibly non-constant
magnetic field}, i.e. $B$ possibly depending on the spatial
coordinates.
\end{enumerate}

\medskip
\noindent
{\bf Short preview of results:}
The main results concerning the representations of the
algebra (\ref{nccr-plane})
and the spectra of Hamiltonian operators
$H( \hat{\vec X}, \hat{\vec P} \, )$
may be summarized as follows.
\begin{itemize}
\item
For $B \neq 1 / \theta$ (e.g. for $B=0$),
there exists a linear {\em invertible} transformation relating
the operators ${\hx}_i , {\hp}_j$
to operators $X_i, P_j$
satisfying canonical commutation relations.
By contrast, if $B$ and $\theta$ are related by $B =1 /\theta$,
the transformation from $\hx_i, \hp_j$ to
 $X_i , P_j$ is no longer invertible.
The singular behavior is also reflected by the fact that the so-called
canonical limit $(\theta, B) \to (0,0)$
does not exist if $B =1 /\theta$.
Moreover, in this case the four-dimensional phase space 
degenerates to a two-dimensional one in the sense 
that for any irreducible representation of the 
commutator algebra, the  representation of the $\hx_i$
alone becomes irreducible. 
\item
While the Jacobi identities for the algebra (\ref{nccr-plane})
involving the constant  magnetic field $B$ are trivially
satisfied, they are violated {\em for a non-constant field} $B$ if
$\theta \neq 0$. Thus, contrary to standard quantum mechanics, 
such a magnetic field forces us to introduce the interaction
into the Hamiltonian by means of a vector potential and
a minimal coupling. The noncommutativity of configuration space 
then leads to a {\em non-Abelian gauge structure} 
and thus to self-interactions of the
gauge potential which do not exist in standard quantum mechanics.
\item
The spectrum of a typical Hamiltonian operator like 
$H( \hat{\vec X}, \hat{\vec P}) = {1 \over 2m} \, \hat{\vec P} ^{2}
+ V(\hx_1, \hx_2)$
depends on the {\em ordering} of the noncommuting operators
${\hx}_1 , {\hx}_2$.
Apart from some special cases, the energy
spectrum for $\theta \neq 0$
differs notably from the one for $\theta =0$.
Non-polynomial potentials $V(\hx_1, \hx_2)$  become {\em non-local}
when expressed in terms of canonical coordinates $X_1, X_2$.
Depending on the chosen parametrization of the phase space variables
$\hx_i, \hp_j$ in terms of canonical variables 
 $X_i, P_j$,
the Hamiltonian may take different disguises.
However, in many instances, the parameter $\theta$ resembles
a constant magnetic field.
\end{itemize}

\medskip
\noindent
{\bf Outline of the presentation:}
To start with, we recall the description of a particle in a
magnetic field within standard quantum mechanics 
since it represents a useful guideline for
noncommutativity in quantum mechanics and a starting point for
generalizations. Section 3 deals with noncommutative classical
mechanics. In section 4, we discuss the representations of the
commutator algebra of NCQM. The remainder of the text is devoted to
the study of physical models of NCQM: we discuss some general 
properties of Hamiltonian operators and of their spectra,
 and then treat in detail 
the problem of a (possibly non-constant) magnetic field.
The text concludes with a summary and some remarks.

\section{Reminder on a particle in a magnetic field
in QM}\label{landausystem}

In non-relativistic quantum mechanics in ${\bf R} ^d$, the
coordinates $X_i$ and momenta $P_j$ satisfy the CCR's.
It is worthwhile recalling that the Schr\"odinger
representation on $L^2 ({\br} ^d, d^dx )$ (i.e. $P_j = {\hbar
\over \ri} {\partial \over
\partial x_j}$ and $X_i \, =$ operator of multiplication by the
real variable $x_i$) is the only possible irreducible realization
of the CCR's
 up to unitary equivalence\footnote{In 
the relation $[ {X}_k , {P}_k ] = \ri \hbar
\Id $ at least one of the self-adjoint operators ${X}_k , {P}_k$
has to be unbounded, henceforth it is not defined on all of
Hilbert space, but only on some subspace. For this reason, one
usually exponentiates all of these operators so as to obtain
unitary (i.e. bounded) operators satisfying Weyl's form of the
CCR's: the cited classification theorem only makes sense for this
form of the CCR's.}~\cite{rs,beh}.

\subsection{General magnetic fields}

Concerning  the algebra of NCQM and its representations,
it is
useful to have in mind the standard quantum mechanical
treatment~\cite{galindo} of a charged particle in three dimensions
which is coupled to a magnetic field $\vec B( \vec x \, )$
deriving from a vector potential $\vec A ( \vec x \, )$. In the
next two paragraphs, we recall the two different, though
equivalent descriptions of the problem under consideration.

\subsubsection{Hamiltonian involving a vector potential}

We start from the coordinates $X_i$
and the components $P_j$ of the canonical momentum
$\vec P$ satisfying the CCR's (\ref{heisen}).
If the spin of the particle is not taken into account,
the interaction
with the magnetic field $\vec B = \vec{{\rm rot}} \, \vec A$
is simply described by means of the so-called minimal coupling,
i.e. by the Hamiltonian
\begin{equation}
\label{mincoup}
H = {1 \over 2m} \, (\vec P - {e \over c} \vec A \, ) ^{2}
\, .
\end{equation}

\subsubsection{Commutation relations involving the magnetic field}

The Hamiltonian (\ref{mincoup})
is quadratic in the components of the kinematical
(gauge invariant) \cite{ctdl} momentum
$\vec{\Pi} \equiv \vec P - {e \over c} \vec A$:
\begin{equation}
H = {1 \over 2m} \, \vec{\Pi}^{\, 2}
\, .
\end{equation}
Thus, it has the same form as the {\em free}
particle Hamiltonian $H_0 \equiv {1 \over 2m} \, \vec{P} ^{\; 2}$
with $\vec P$ replaced by $\vec \Pi$.
The momentum variables $\Pi _i$
do not commute with each other,
rather they satisfy the {\em non-canonical} commutation relations
\begin{eqnarray}
\label{veloc}
{[ \Pi_i , \Pi_j ]}
\!\!\! &=& \!\!\!
\ri \hbar \, {e \over c} \, B_{ij}
\, , \qquad {\rm where} \ \;
B_{ij} \equiv \partial_i A_j -\partial_j A_i
= \varepsilon_{ijk} B_k
\, ,
\\
{[ X_i , X_j ]} \!\!\! &=& \!\!\! 0
\, , \qquad
{[ {X}_i , {\Pi}_j ]}  =  \ri \hbar \, \delta_{ij} \mathds{1}
\qquad \ \ {\rm for} \ \; i,j \in \{ 1,2,3 \}
\, .
\nonumber
\end{eqnarray}
In other words, when starting from the description of
a free particle,
the coupling of the particle to a magnetic field
can simply be described by replacing the canonical
momentum $\vec P$ in the free Hamiltonian by the
kinematical momentum $\vec \Pi$ whose components
have a non-vanishing commutator: it is the magnetic field
which measures this {\em noncommutativity}.

We note that the Jacobi identities for the quantum algebra
(\ref{veloc}) are satisfied
for any field strength $\vec B( \vec x \, )$
and,  in particular, one has
$[X_i, [ \Pi_j, \Pi _k ]] + \mbox{circular permutations} =0$.
For $\vec  B \neq \vec 0$, the transformation
$(\vec X , \vec P \, ) \mapsto (\vec X , \vec \Pi \, )$
is {\em not} unitary since it modifies the commutators (\ref{heisen}).

\subsection{Constant magnetic field (Landau system)}

In the particular case where the magnetic field is {\bf constant},
the coordinate axes can be oriented such that $\vec B = B \vec
e_3$. The three-dimensional 
problem then reduces to a two-dimensional one, namely the problem of 
a particle in the
$x_1 x_2$-plane subject to a vector potential $\vec A = (A_1, A_2) $ such
that $B=
\partial_1 A_2 - \partial_2 A_1$: the
Hamiltonian then reads as
\begin{equation}
\label{hemf}
H= {1 \over 2m} \, ( \Pi_1^2 +  \Pi_2^2 )
\, ,
\end{equation}
if one discards the term ${1 \over 2m} \, P_3^2$ describing a free
particle motion along the $x_3$-direction. The present physical
system (i.e. a charged particle in a plane subject to a constant
magnetic field that is perpendicular to this plane) is often
referred to as the {\bf Landau system}. For this problem,
convenient choices of the vector potential are the {\em symmetric
gauge} $(A_1, A_2 ) = (-{B \over 2} x_2  ,  {B \over 2} x_1)$
and 
the {\em Landau gauge}, i.e. $(A_1, A_2 ) = ( 0, B x_1)$ or $(A_1,
A_2 ) = ( -B x_2, 0)$.

The energy spectrum of the Landau system
can be obtained by a simple argument~\cite{dunne}.
In fact, the Hamiltonian only depends on
$\vec \Pi = (\Pi_1 , \Pi_2 )$ and the latter variables satisfy the commutation
relation $[\Pi _1 , \Pi_2 ] = \ri \hbar \, {e \over c} \, B \Id$, or equivalently
\begin{equation}
\label{rewrite}
[Q , \Pi_2 ] = \ri \hbar \, \Id
\qquad {\rm with} \quad
Q \equiv \ds{c \over eB} \, \Pi_1
\, .
\end{equation}
Thus, the operators
$Q$ and $\Pi_2$ can be viewed as canonically conjugate variables
in terms of which the Hamiltonian reads as
\begin{equation}
\label{hamos}
H = \ds{1 \over 2m} \, \Pi_2^2 + \ds{1 \over 2 } \, m \omega _B ^2 Q^2
\qquad {\rm with} \quad \omega _B \equiv \ds{|eB| \over mc }
\, .
\end{equation}
Accordingly, the Landau system is equivalent to a linear harmonic oscillator
and its eigenvalues (the so-called Landau levels)
are given by
\begin{equation}
\label{speclandau}
E_n = \hbar \, \ds{|eB| \over mc } \, (n + {1 \over 2} )
\qquad {\rm with} \quad n=0,1,2,\dots
\end{equation}
Each of these energy levels is infinitely degenerate
(with respect to
the momentum eigenvalues
$p_1 \in {\bf R}$) which reflects the fact that the Landau system
is actually two-dimensional.

\section{Noncommutative classical mechanics}\label{classlim}

Instead of an algebra of commutators, one can consider its
classical analogon~\cite{aca1, duval} involving Poisson brackets
$\{ \cdot , \cdot \}_{{\rm PB}}$ of functions depending on the
real variables $x_1, x_2, p_1, p_2$.

\subsection{Poisson algebra and its representations}

\subsubsection{Poisson brackets}

The classical algebra
 associated to
 (\ref{nccr-plane}) reads as~\cite{aca1, aca3}
\begin{equation}
   \label{spb}
\{ x_1, x_2 \} _{{\rm PB}} = {\theta}
\, , \qquad
\{ p_1, p_2 \} _{{\rm PB}} = {B}
\, , \qquad
\{ x_i, p_j \} _{{\rm PB}} = {\delta_{ij}}
\, .
\end{equation}
The so-called ``exotic'' algebra~\cite{duval,duvalhor,horvathy}
amounts to modifying these brackets by a factor
$1/ \kappa$ where
\begin{equation}
\label{coeffkappa}
\kappa \equiv 1 - B \theta
\, ,
\end{equation}
i.e.
\begin{equation}
\label{exob}
\{ x_1, x_2 \} _{{\rm PB}} = {\theta \over \kappa}
\, , \qquad
\{ p_1, p_2 \} _{{\rm PB}} = {B \over \kappa}
\, , \qquad
\{ x_i, p_j \} _{{\rm PB}} = {\delta_{ij} \over \kappa}
\, .
\end{equation}
The expression $\kappa$ is a characteristic parameter
for the algebraic system under study
and will frequently reappear in the sequel.

If we denote the phase space coordinates collectively by
$({\xi} ^I)_{I= 1,...,4} \equiv (x_1 , x_2, p_1, p_2)$,
the {\em Poisson bracket of two functions} $F$ and $G$ on phase space
reads as
$\{ F , G \}_{{\rm PB}} \equiv \sum_{I,J =1}^4 \Omega^{IJ} \,
{\partial F \over \partial {\xi} ^I } \,
{\partial G \over \partial {\xi} ^J }$,
so that
$\{ \xi^I , \xi^J \}_{{\rm PB}} = \Omega^{IJ}$,
and the symplectic $2$-form is given by
\begin{equation}
\label{symform}
\omega = {1 \over 2} \,
\sum_{I,J =1}^4 \,
\omega _{IJ}
\, d{\xi} ^I
\wedge d{\xi} ^J
\, ,
\end{equation}
with $ \omega _{IJ} \equiv (\Omega ^{-1} ) _{IJ}$.
The symplectic matrix $\Omega \equiv ( \Omega^{IJ} )$
for the standard and exotic algebras (\ref{spb}) and  (\ref{exob})
is respectively given by
\begin{equation}
\label{symmat}
\Omega_{{\rm standard}} =
\left[
\begin{array}{cccc}
0 & \theta & 1 & 0 \\
-\theta & 0 & 0 & 1 \\
-1  & 0 & 0 & B \\
0 & -1  & -B & 0
\end{array}
\right]
\, , \qquad
\Omega_{{\rm exotic}} = \ds{1 \over \kappa} \, \Omega_{{\rm standard}}
\, .
\end{equation}
Note that ${\rm det} \, \Omega_{{\rm standard}}
= \kappa^2$, hence ${\rm det} \, \Omega_{{\rm exotic}}
= \kappa^{-2}$.
For $\theta =0$ or  $B=0$,
the matrices $\Omega_{{\rm standard}}$ and $\Omega_{{\rm exotic}}$
coincide with each other.
The brackets (\ref{exob})
defining the exotic algebra
diverge for $\kappa \to 0$,
but not so the associated symplectic form.
For the ``standard'' algebra (\ref{spb}),
one has the reversed situation,
i.e. one does not have a well defined symplectic form.

\subsubsection{Transformation to canonical coordinates}\label{darboux}

By definition, the phase space ${\bf R} ^{4}$ equipped with a
non-degenerate, antisymmetric bilinear form (\ref{symform})
is a {\em symplectic vector space}~\cite{mars,arnold}.
(The condition of non-degeneracy
is equivalent to ${\rm det} \, (\omega_{IJ}) \neq 0$.)
For such a space, one can find a change of basis
(i.e. an {\em invertible} linear transformation
of coordinates, $(\xi ^I ) \mapsto (\xi ^{\prime I})$)
such that the matrix $ (\omega_{IJ})$ goes over into the canonical
(normal) form
$J = \left[
\begin{array}{cc}
0 & \Id _2 \\
-\Id _2 & 0
\end{array}
\right]$. This result~\cite{mars,arnold}, which is also referred to
as the {\bf linear Darboux theorem}, represents a special case of
the Darboux theorem which applies to general phase space 
manifolds.
 The explicit form of the linear
transformation $(\xi ^I) \mapsto (\xi ^{\prime I})$ for the
algebra (\ref{nccr-plane}) will be discussed in section 4 in the
context of quantum mechanics. Here, we only emphasize that this
transformation is not canonical (in the sense of Hamiltonian
mechanics) since it modifies the Poisson brackets.

\subsubsection{Case of non-constant magnetic field}

Let us consider the Poisson algebra (\ref{spb})
with constant parameter $\theta \neq 0$
and with a  parameter  $B$ which depends on $\vec x$
(but not on $\vec p \, $). Then, 
\[
\{ x_i , \{ p_1 , p_2 \} \}_{{\rm PB}} + \mbox{circular permutations}
\, = \, \theta \varepsilon_{ij} \, \ds{\partial B \over \partial x_j}
\qquad {\rm for} \ \; i=1,2
\, ,
\]
i.e. the Jacobi  identities are only satisfied if 
$B$ is a constant magnetic field.
By contrast, the exotic algebra (\ref{exob})
with constant parameter $\theta \neq 0$
allows for a ${\vec x}$-dependent $B$-field
since such a field is compatible with the
corresponding Jacobi identities~\cite{horvathy}. 
Thus, in the noncommutative plane, the coupling of a particle
to a {\em non}-constant magnetic field can be achieved
by the exotic approach, but not by the standard one.
For $\theta =0$, the two approaches coincide with each other.
We will come back to the exotic algebra
in section~\ref{exotic}.

\subsection{Dynamics}

The time evolution of a function $F$ on phase space 
 is given by
{\em Hamilton's equation of motion}
$\dot{F} = \{ F, H \}_{{\rm PB}}$
where $H$ is the Hamiltonian function.
We will only consider the case of a constant magnetic field.

\subsubsection{Poisson brackets involving the magnetic field}

The dynamics based on the Poisson algebra (\ref{spb})
has been studied in references~\cite{aca1,aca3,aca4}.
For the Hamiltonian
$H= {1 \over 2m} \, \vec p ^{\; 2} + V(x_1, x_2) $,
one finds the equation of motion
\begin{equation}
\label{eommf}
m \ddot{x} _i
= - \kappa \, {\partial V \over \partial x_i}
+ B \varepsilon_{ij} \dot{x} _j
+ m \theta \varepsilon_{ij}
\, {d \over dt} \left( {\partial V \over \partial x_j } \right)
\qquad {\rm for} \ \; i \in \{ 1,2 \}
\, ,
\end{equation}
where $\kappa  \equiv 1 - B \theta$.
If the singular limit $\kappa  \to 0$ is taken,
the first (Newton-like) force on the right-hand-side
vanishes and the number of dynamical degrees of freedom
is reduced by half.
(This result is related to the Peierls substitution
discussed in the appendix.)

Some specific potentials, e.g. the one of the
anisotropic harmonic oscillator,
are elaborated upon in reference~\cite{aca3}.
Here, we will only discuss the case $V =0$,
i.e. the dynamics of a charged particle (of unit charge)
in the noncommutative plane subject to a constant magnetic
field which is perpendicular to this plane.
The equation of motion (\ref{eommf}) then reduces to
\begin{equation}
\label{eommag}
m \ddot{x} _i = B \varepsilon_{ij} \dot{x} _j
\qquad {\rm for} \ \; i \in \{ 1,2 \}
\, .
\end{equation}
Thus, we have exactly the same  equation of motion
as for $\theta =0$.
Indeed, the fact that $H$ does not depend on the
coordinates $x_i$ implies
that the noncommutativity of these coordinates
does not manifest itself in the classical trajectories,
though it affects some
physical quantities (like the density of states per unit area 
-- see section 4 below for the quantum theory).

\subsubsection{Hamiltonian involving a vector potential}\label{magnet}

Let us now consider the Poisson algebra (\ref{spb}) with $B=0$ and
introduce a constant magnetic field ${\cal B}$
deriving from a vector potential
$\vec {\cal A} = ({\cal A}_1, {\cal A}_2)$, i.e.
$\partial_1 {\cal A}_2 
- \partial_2 {\cal A}_1 = {\cal B}$.
The equations of motion following from the minimally coupled
Hamiltonian $H= {1 \over 2m} ( \vec p  - \vec {\cal A} \, )^2$
then read as~\cite{aca1}
\begin{equation}
m \dot{x}_i =
(p_k - {\cal A}_k ) \left( \delta_{ik} - \theta \varepsilon_{ij}
\, \partial_j {\cal A}_k  \right)
\, , \qquad
m \dot{p}_i =
(p_k - {\cal A}_k ) 
\, \partial_i {\cal A}_k
\, .
\end{equation}
In the symmetric gauge
${\cal A}_i = -{{\cal B} \over 2} \varepsilon_{ij} x_j$,
these equations yield
\begin{equation}
\label{classsg}
m \ddot{x} _i =  \varepsilon _{ij} \dot{x}_j
\, {\cal B} \, (1 + {1 \over 4} \theta {\cal B} )
\end{equation}
and in the Landau gauge $({\cal A}_1,{\cal A}_2 ) = (0, {\cal B} x_1)$
we get
\begin{equation}
\label{classlg}
m \ddot{x} _1 = \dot{x}_2
\, {{\cal B} \over 1 + \theta {\cal B} }
\, ,
\qquad
m \ddot{x} _2 = -\dot{x}_1
\, {\cal B} \, (1 + \theta {\cal B} )
\, .
\end{equation}
Thus, it seems that
for  $\theta \neq 0$ the equations of motion (resulting from the
 minimally coupled
Hamiltonian $H= {1 \over 2m} ( \vec p  - \vec {\cal A} \, )^2$)
depend on the chosen gauge, 
and that they differ from those obtained in equation (\ref{eommag})
(by introducing the magnetic field into the Poisson brackets). 
This point can be elucidated by evaluating the Poisson bracket 
between the components of the kinematical momentum 
$\vec p - \vec {\cal A}$:
\begin{equation}
\{ p_1 - {\cal A}_1 , p_2 - {\cal A}_2 \}_{{\rm PB}} =
\partial_1 {\cal A}_2 
- \partial_2 {\cal A}_1
+ \{ {\cal A}_1 , {\cal A}_2 \}_{{\rm PB}} 
\equiv {\cal F}_{12} 
\, .
\end{equation}
The resulting expression for the field strength ${\cal F}_{12} $
involves a quadratic term $\{ {\cal A}_1 , {\cal A}_2 \}_{{\rm PB}} $
which is characteristic for non-Abelian gauge field theories.
The field ${\cal F}_{12} $ appears in the equation of motion 
for the velocity $v_i \equiv \dot{x}_i$: indeed, 
by differentiating equations (\ref{classsg}) and (\ref{classlg}), 
we obtain the equation of motion 
$\ddot{v}_i + \omega^2 v_i =0$ where the frequency
$\omega$ is defined by 
\[
\omega \equiv {|{\cal F}_{12}|  \over m} 
\, , \quad {\rm with} \ \  
{\cal F}_{12} = 
\left\{ 
\begin{array}{ll}
{\cal B} \, (1 + {1 \over 4} \theta {\cal B} ) & 
\mbox{for the symmetric gauge} \\
{\cal B}  & 
\mbox{for the Landau gauge} \, .
\end{array}
\right.
\]
Thus, the gauge potentials 
${\cal A}_i = -{{\cal B} \over 2} \varepsilon_{ij} x_j$ 
and 
$({\cal A}_1,{\cal A}_2 ) = (0, {\cal B} x_1)$
do not define the same noncommutative field strength ${\cal F}_{12} $.
We will encounter a completely analogous situation in NCQM
(sections \ref{ahat} and \ref{starpro}) and we will discuss in that context 
the non-Abelian gauge transformations, the introduction of a 
constant $\theta$-independent  field strength and the comparison 
between the different approaches to such a magnetic field.

\section{NCQM: representations of the algebra}\label{algncqm}

{\em In the sequel of the text, we generally consider a charge $e=1$ and a system of
units such that $\hbar \equiv 1 \equiv c$.} {\em We always assume
that the canonical operators $X_i$ and $P_j$ operate on $L^2
({\br}^2 , dx_1 dx_2)$ as in the standard Schr\"odinger
representation,} i.e. $X_i$ acts as multiplication by $x_i$ and
$P_j = {\hbar \over \ri} \, \partial _j$.

\subsection{Representations of the algebra}

If the
{\bf canonical limit} (or so-called commutative limit)
 $(\theta, B) \to 0$
exists and is considered,
the quantum algebra (\ref{nccr-plane})
reduces to the Heisenberg algebra (\ref{heisen})
so that the operators $\hx_i,  \hp_j$
should reduce to the operators
$X_i, P_j$ satisfying (\ref{heisen}).
Thus, it is natural to look for representations of the algebra
(\ref{nccr-plane}) for which the operators
$\hx_i, \hp_j$
are operator-valued functions of the real
parameters $\theta, \, B$
and of the operators $X_i, P_j$.

To determine such representations,
one generally considers a linear transformation
$(\vec X , \vec P \, ) \mapsto (\hat{\vec X} , \hat{\vec P} \, )$
whose transformation matrix depends on the
noncommutativity parameters, e.g. see references~\cite{smail, li}.
In the following, we discuss different representations
of (\ref{nccr-plane}) and in
section~\ref{fake} we will come back to the existence of the
canonical limit.

\medskip

\paragraph{(i) \underline{Case $\theta =0$} :}

\smallskip

A simple particular case of (\ref{nccr-plane})
 is given by the Landau system discussed
in section~\ref{landausystem}:
then, we have the algebra
\begin{equation}
\label{landau}
[ \hx _1 , \hx _2 ] = 0
\, , \qquad
[ {\hp}_1 , {\hp}_2 ] = \ri  B \Id
\, , \qquad
[ {\hx}_i , {\hp}_j ] = \ri \delta_{ij} \Id
\end{equation}
and different gauge choices
$( A_1 ( \vec x \, ) , A_2 ( \vec x \, ))$ with
$\partial_1 A_2 - \partial_2 A_1 = B$
provide different representations
of the algebra (\ref{landau}), 
all of which have the form
\begin{equation}
\label{repmag}
\hx _i = X_i
\, , \qquad
\hp _j = \Pi_j \equiv P_j -  A_j( \vec X \, )
\quad {\rm with} \ \; \partial_1 A_2 - \partial_2 A_1 = B
\, .
\end{equation}
Vector potentials $\vec A, \vec A ^{\prime}$
giving rise to the same magnetic field $B$
are related by a gauge transformation
$\vec A \mapsto \vec A ^{\prime} = \vec A
+ \vec{\nabla} \alpha (\vec X \, )$
which goes along with 
a unitary transformation of operators (and states):
$\hx_i ^{\prime} = U \hx_i U^{-1}$ and
$\hp_j ^{\, \prime} = U \hp_j U^{-1}$ with $U= \re^{\ri \alpha}$.
In fact~\cite{gruemm}, all representations of (\ref{landau})
are unitarily equivalent\footnote{For the reasons 
indicated in the footnote to section 2, 
one should exponentiate the unbounded self-adjoint 
operators $\hx_i, \hp_j$
so as to obtain Weyl's form of commutation relations: 
the classification theorem for the representations 
actually refers to this form of the
commutation relations.} to the representation (\ref{repmag}).

\medskip

\paragraph{(ii) \underline{Case $B=0$} :}

\smallskip

A completely analogous situation is the case
where the spatial coordinates
do not commute whereas the momenta
commute:
\begin{equation}
\label{nctheta}
[ \hx _1 , \hx _2 ] = \ri \theta \Id
\, , \qquad
[ {\hp}_1 , {\hp}_2 ] = 0
\, , \qquad
[ {\hx}_i , {\hp}_j ] = \ri \delta_{ij} \Id
\end{equation}
Then,
\begin{equation}
\label{repthe}
\hp _i = P_i
\, , \qquad
\hx _j = X_j - \tilde A _j (\vec P \, )
\quad \ {\rm with} \ \  \;
\ds{\partial \tilde A_1 \over \partial P_2}
- \ds{\partial \tilde A_2 \over \partial P_1} =\theta
\end{equation}
provides a representation of the algebra (\ref{nctheta}).
In fact, in the present context it is appropriate to consider
Schr\"odinger's {\em momentum space} representation for
the canonical operators $P_i$ and $X_j$: the operators
$\hp_i$ and $\hx_j$ then act on wave functions
 $\tilde{\psi} \in L^2 ({\br}^2 , dp_1 dp_2)$
according to
\begin{equation}
\hp_i \tilde{\psi} =
p_i \tilde{\psi}
\, , \qquad
 \hx_j \tilde{\psi} =
\ri \, {\partial  \tilde{\psi} \over \partial p_j} -
 \tilde A_j \tilde{\psi}
\, .
\label{msr}
\end{equation}
The parameter $\theta$ then measures the  noncommutativity
of the $\vec p$-space covariant derivatives.
Obviously, the vector field $(\tilde A_1, \tilde A_2)$
in the ${\vec p}$-plane
is analogous to a vector potential
in the ${\vec x}$-plane,  and $\theta$ to
a magnetic field.
For a given $\theta$,
different choices of the vector field $(\tilde A_1, \tilde A_2)$
may be viewed as different {\em ``gauges''} for
$\hat{\vec X}$
(just as different choices of the vector potential
$(A_1, A_2)$ define different gauges for the covariant derivatives
$\hp _j$ in equation (\ref{repmag})).
A change of gauge,
$\tilde A_j \mapsto \tilde A_j ^{\prime} =  \tilde A_j
 + {\partial \tilde{\alpha} \over \partial p_j} (\vec p \, )$
induces a unitary
transformation of $\hx_i$ and $\hp_j$
by means of 
the operator $V= \re^{\ri \tilde{\alpha}(\vec P \, )}$.
Analogously to case (i), all representations of (\ref{nctheta})
are unitarily equivalent to the representation (\ref{msr}).

\paragraph{(iii) \underline{General case} :}

\smallskip

If we combine the noncommutativities (\ref{landau}) and (\ref{nctheta}),
we obtain the algebra
\begin{equation}
\label{thb}
[ \hx _1 , \hx _2 ] =  \ri \theta \Id
\, , \qquad
[ {\hp}_1 , {\hp}_2 ] = \ri  B \Id
\, , \qquad
[ {\hx}_i , {\hp}_j ] = \ri \delta_{ij} \Id
\, ,
\end{equation}
which can be represented by combining particular
realizations of
(\ref{landau}) and (\ref{nctheta}), e.g.~\cite{nair}
 \begin{equation}
\begin{array}{lcl}
\hx _1 = X_1
& \qquad &
\hp _1 = P_1 + BX_2
\\
\hx _2 = X_2 + \theta P_1
& \qquad &
\hp _2 = P_2
\, .
\end{array}
\label{thbt}
\end{equation}
While the `minimal' representation (\ref{thbt})
amounts to considering the {\em ``Landau gauge''} for $\hat{\vec X}$,
we can also choose the {\em ``symmetric gauge''} for $\hat{\vec X}$
if $\theta \neq 0$~\cite{smail}:
\begin{equation}
\begin{array}{lcl}
\hx _1 = a X_1  - \ds{\theta \over 2a} \, P_2
& \qquad &
\hp _1 = c P_1 + d X_2
\\
\hx _2 = a X_2 +  \ds{\theta \over 2a}  \, P_1
& \qquad &
\hp _2 = c P_2 - d X_1
\, ,
\end{array}
\label{euro}
\end{equation}
where
\begin{equation}
a \in {\bf R}^{\ast}
\, , \qquad
c = \ds{1 \over 2a} \, ( 1 \pm \sqrt{\kappa} \, )
\, , \qquad
d = \ds{a \over \theta} \, ( 1 \mp \sqrt{\kappa} \, )
\, ,
\label{coeffcd}
\end{equation}
with $\kappa$ as defined in (\ref{coeffkappa}): $\kappa = 1 - B \theta$.
The expressions (\ref{thbt}) and (\ref{euro})
specifying
$(\hat{\vec X}, \hat{\vec P}\, )$
in terms of
$(\vec X , \vec P \, )$
are invertible
if and only if the
functional determinant
\begin{equation}
\label{jacobi}
{\rm det} \, \left[ \ds{\partial (\hat{\vec X}, \hat{\vec P}\, )
\over \partial  ( {\vec X},  {\vec P}\, )} \right]
= \kappa
\end{equation}
does not vanish, i.e. if and only if  $B\neq 1 / \theta$.
As we noticed in section \ref{darboux}, this result reflects
the linear Darboux theorem which holds in classical mechanics
if the symplectic matrix defining the Poisson brackets
of coordinates and momenta is invertible.

For $B = 1 / \theta$,
the canonical limit $(\theta , B) \to 0$ does not exist for the
algebra (\ref{thb}) nor for the given representations:
we will come back to this case in section~\ref{fake}.
At this point, we only note that the representation  (\ref{thbt})
is reducible for $B = 1 / \theta$ since the operators
(\ref{thbt}) then leave invariant the linear space of functions
of the form $\psi (x_1, x_2) = f(x_1)
\; {\rm exp} \, [ \ri ( \lambda -  {x_1 \over \theta } ) x_2 ]$.

If the coordinates and momenta are gathered
into a phase space vector
\[
\hat{\vec u} \equiv \left( \hat u _I  \right)_{I = 1, \dots, 4}^t =
\left( {\hx}_1, {\hp}_1, {\hx}_2, {\hp}_2 \right) ^t
\, , 
\]
then the commutation relations (\ref{thb}) take the form 
$[ \hat u _I, \hat u_J ] = \ri \hat M _{IJ} \Id$  
where $\hat M$ is a constant antisymmetric $4\times 4$ matrix.
As discussed in references~\cite{hatzi,jonke}, 
the matrix $\hat M$ can be block-diagonalized by means 
of an $O(4)$ matrix $R$ (i.e. $R^t R = \Id$). 
For $B \neq 1/\theta$, the components of 
the transformed phase space vector ${\vec u} = R^t \hat{\vec u}$
can be rescaled so as to obtain 
new phase space variables
$X_1 , P_1,X_2 , P_2$ which satisfy CCR's.
Thus, for variables $(\hat{\vec X}, \hat{\vec P}\, )$
satisfying the algebra  (\ref{thb}) with  $B \neq 1/\theta$,
there exists an invertible linear transformation to 
canonical variables $( {\vec X},  {\vec P}\, )$. 
Since all representations of the CCR's are unitarily equivalent, 
this also holds for the representations of  the algebra  (\ref{thb})
if $B \neq 1/\theta$.

To conclude, we note that the 
algebra (\ref{thb})
can be decoupled by virtue of appropriate linear combinations
of generators~\cite{nair, bellucci}: for $\theta \neq 0$,
we may consider the change of generators
$(\hat{\vec X}, \hat{\vec P} \, )
\mapsto (\hat{\vec X}, \hat{\vec K}\, )$ given by
\begin{equation}
\label{chof}
\hat K_j = \hat P_j - {1 \over \theta} \, \varepsilon _{jk} \hat X_k
\qquad  {\rm for} \ \; j =1,2
\, ,
\end{equation}
which yields a direct sum of two algebras of the same form:
\begin{equation}
\label{decoup}
[ \hx _1 , \hx _2 ] =  \ri \theta \Id
\, , \qquad
[ {\hat K}_1 , {\hat K}_2 ] = -\ri  \, \ds{\kappa \over \theta} \, \Id
\, , \qquad
[ {\hx}_i , {\hat K}_j ] = 0
\, .
\end{equation}
If
$B= 1/\theta$ (i.e. $\kappa =0$), one concludes from (\ref{decoup})
that $\hat K_1$ and $\hat K_2$ commute with all generators
of the algebra (i.e. they belong to the center of the algebra
and represent Casimir operators).
By Schur's lemma, they are $c$-number operators for any irreducible
representation, i.e. $\hat K_i = \lambda _i \Id$
with $ \lambda _i \in {\bf R}$.
In this case,
the only nontrivial commutator is the one involving
$\hx _1$ and $\hx _2$ which may be rewritten as
\begin{equation}
\label{xnt}
 [ \hx _1 , \hat{\Pi} ] =  \ri \Id
\qquad {\rm with} \ \;
\hat{\Pi} = {1 \over \theta} \,  \hx _2
\, ,
\end{equation}
i.e. we have canonically conjugate variables in one dimension.

\subsection{About the relation between the parameters $\theta$ 
and $B$}\label{fake}

Let us start from noncommuting coordinates in the plane
and assume as before 
that the commutator between $\hat{\vec{X} }$ and $\hat{\vec{P} }$
is canonical~\cite{sochi}:
\begin{equation}
[ \hx _1 , \hx _2 ] =  \ri \theta \Id
\, , \qquad
[ {\hx}_i , {\hp}_j ] = \ri \delta_{ij} \Id
\, , \qquad
[ {\hp}_1 , {\hp}_2 ]
\, \propto \, \ri \Id
\, .
\end{equation}
Furthermore, let us assume that the considered representation
of this algebra is {\em irreducible}.

For the linear combination
$\hat K_j = \hat P_j - {1 \over \theta} \, \varepsilon _{jk} \hat X_k$
introduced in equation (\ref{chof}), we have
\begin{eqnarray}
\nonumber
[ {\hat K} _i , {\hat X} _j ]
\!\!\! & = & \!\!\! 0
\, , \qquad
{[ {\hat K} _1 , {\hat P} _1 ]}
= 0 =
{[ {\hat K} _2 , {\hat P} _2 ]}
\\
{[ {\hat K} _1 , {\hat P} _2 ]}
\!\!\! & = & \!\!\!
[ {\hat P} _1 , {\hat P} _2 ]
-\ri  \, \ds{1 \over \theta} \, \Id
= - [ {\hat K}_2 , {\hat P} _1 ]
\label{schu}
\end{eqnarray}
and $[ {\hat K} _1 , {\hat K} _2 ] = [ {\hat P} _1 , {\hat P} _2 ]
-\ri  \, \ds{1 \over \theta} \, \Id$.
Thus, there are two particular cases:
\[
[ {\hp}_1 , {\hp}_2 ] = \ri  \, \ds{1 \over \theta} \, \Id
 \qquad \quad {\rm or}  \quad \qquad
[ {\hp}_1 , {\hp}_2 ] = \ri  \, B \, \Id
\quad {\rm with} \ \; B\neq \ds{1 \over \theta}
\, .
\]
For the first case, it follows from equations (\ref{schu})
that the operators
${\hat K}_1$ and ${\hat K}_2$ commute with
$\hat{\vec X}$ and $\hat{\vec P}$. By Schur's lemma
they are constant multiples of the identity operator so that
$\hat{\vec P}$ is a function of $\hat{\vec X}$, e.g.
for vanishing constants:
$\hat P_j = {1 \over \theta} \, \varepsilon _{jk} \hat X_k$.
(For instance, 
for the irreducible representation (\ref{euro})
we have $\hat K_1 = 0 = \hat K_2$
if $B \equiv 1 / \theta$.)
Henceforth, the only independent commutator is then given
by  $[ \hx _1 , \hx _2 ] =  \ri \theta \Id $
and describes a $1$-dimensional canonical system:
we have a degeneracy of the representation of the
quantum algebra
in the sense that the representation of the algebra of the $\hx_i$
alone becomes irreducible~\cite{bellucci, horvathy}.
In view of this result, the phase space with $B = 1/\theta$ 
might be referred to as 
a ``degenerate'' or ``singular'' noncommutative space.
For this space, the canonical limit
$\theta \to 0$ does not exist and 
one cannot express the
variables $\hat{\vec X}$ and $\hat{\vec P}$
in terms of canonical variables ${\vec X}$ and ${\vec P}$
by virtue of an {\em invertible} transformation
-- see equation (\ref{jacobi}). 

By contrast, in the second case
the canonical limit
$(\theta , B) \to 0$ can be taken and 
there exists an {\em invertible} transformation
relating the
variables $( \hat{\vec X}, \hat{\vec P} \, )$
and $({\vec X}, {\vec P} \, )$,
e.g. the transformation (\ref{thbt}).
Particular instances of this case 
are given by $\theta =0, \, B\neq 0$
and   $\theta \neq 0, \, B=0$.

\section{NCQM systems: generalities}

\subsection{About Hamiltonians}

A given Hamiltonian $H( {\vec X} , {\vec P} \, )$ of standard
quantum mechanics in ${\bf R} ^d$ can be generalized to a
Hamiltonian $H( \hat{\vec X} ,   \hat{\vec P} \, )$ of NCQM by
replacing the canonical coordinates and momenta by non-canonical
ones. (The ordering problem for the variables $\hx_1, \dots ,
\hx_d$ is to be discussed in the next subsection.) The so-obtained
Hamiltonian is still assumed to act on standard wave functions
$\psi \in L^2 ({\bf R} ^d, d^dx)$. This action is clearly defined  
once an explicit representation of $\hat{\vec X}$ and $\hat{\vec
P}$ in terms of ${\vec X}$ and ${\vec P}$ is given. 
The independence of the spectral properties of 
$H( \hat{\vec X} , \hat{\vec P} \, )$ from the representation
chosen for $\hat{\vec X} , \hat{\vec P}$ will be addressed 
in section 5.3 below. 

The spectrum of
$H( \hat{\vec X} ,   \hat{\vec P} \, )$ can be determined
perturbatively by treating  the noncommutativity parameters
$\theta _{ij}$ as small perturbative parameters.
In some particular
cases, the spectrum of $H( \hat{\vec X} , \hat{\vec P} \, )$ can
be determined exactly by algebraic methods without referring to any
explicit  representation of $(\hat{\vec X}, \hat{\vec P} \, )$ in
terms of $({\vec X},{\vec P}\, )$.

\subsection{Remarks on the ordering problem}\label{orderprob}

In two-dimensional NCQM with position operators satisfying
$[ \hx_1 , \hx_2 ] = \ri \theta \Id$,
one can consider different ordering prescriptions to define 
an operator  $V(\hx_1, \hx_2)$ associated to a classical
function $V(x_1, x_2)$, 
the most popular one being the {\em Weyl ordering}, i.e. 
the symmetrization  prescription.  
After introducing the operator
$\ha = {1 \over \sqrt{2 \theta}} \, (\hx_1 + \ri \hx_2 )$
and its adjoint which operators 
satisfy $[\ha, \ha^{\dagger} ] = \Id$, 
one can also define the 
{\em normal ordering}
(all operators $\ha$ to the right) or the {\em anti-normal ordering}
(all operators $\ha^{\dagger}$ to the right)~\cite{dunne,horvathy}. 
For instance,
for a central potential given by a power law, one has
at the classical level
\[
V( {\vec x}^{\, 2} \, ) \equiv {\vec x} ^{\, 2N} =
\left( x_1^2 + x_2^2 \right) ^N
\sim (2 \theta )^N
\left( \ha^{\dagger} \ha \right) ^N
\qquad (\, N =1,2, \dots \, )
\, ,
\]
where the tilde symbol indicates that some ordering prescription
needs to be specified for the last expression
at the quantum level. E.g.
for the so-called holomorphic polarization
\`a la Bargmann-Fock~\cite{dunne,horvathy},
one chooses the anti-normal order:
\[
\left( \ha^{\dagger} \ha \right) ^N _{\mbox{anti-normal order}}
\equiv
\ha ^N (\ha ^{\dagger} ) ^N =
( \ha ^{\dagger} a + \Id ) \, (\ha ^{\dagger} \ha + 2\Id )
\cdots  (\ha ^{\dagger} \ha + N \Id )
\, .
\]
Alternatively, the operator-valued function
$V( {\hat r}^2)$ (with $ {\hat r}^2 \equiv \hat{\vec X} ^{\, 2}$),
which is radially symmetric, may
simply be assumed to act on Hilbert space
as multiplication by the classical
function $V(\vec x ^{\, 2})$.

For observables which may be probed experimentally,
one can (and one should) resort to the experimental results
for choosing the ``good'' quantization scheme~\cite{klauder, gazeau}.
However in the case where no precise experimental data are available
(as it appears to be 
 the case for the models of NCQM),
it is in general impossible to find a preferred
quantization scheme (although one may invoke some
extra criteria favoring certain choices~\cite{gerst}).

\subsection{Properties of energy spectra}

Before looking at some specific examples, it is worthwhile
to determine some general properties of the spectrum
of a given Hamiltonian $H( \hat{\vec X} , \hat{\vec P} \, )$.

As emphasized earlier, the reparametrization of
$( \hat{\vec X} , \hat{\vec P} \, )$ in terms of
canonical variables $( {\vec X} , {\vec P} \, )$
does not represent a unitary transformation
since it modifies the commutators.
Henceforth the Hamiltonian  $H( \hat{\vec X} , \hat{\vec P} \, )$
is {\em not} unitarily equivalent to its commutative counterpart
$H({\vec X} , {\vec P} \, )$.
By way of consequence,
their spectra are in general different.

As argued in section 4.1, all representations of the quantum algebra
(\ref{thb}) are unitarily equivalent for $B \neq 1/\theta$, 
i.e. for any two representations
$( \hat{\vec X} , \hat{\vec P} \, )$
and $( \hat{\vec X} ^{\, \prime} , \hat{\vec P}^{\, \prime} \, )$,
we have
$\hx_i ^{\prime} = U \hx_i U^{-1}, \,
\hp_j ^{\prime} = U \hp_j U^{-1}$ with $U$ unitary.
Thus, 
the corresponding Hamiltonians are also unitarily equivalent, 
\begin{equation}
H ( \hat{\vec X} ^{\, \prime} , \hat{\vec P}^{\, \prime} \, )
= U H( \hat{\vec X} , \hat{\vec P} \, ) \, U^{-1}
\, , 
\end{equation}
which ensures that
the energy spectrum does not depend on the representation
that has been chosen for the quantum algebra.

By way of example, we briefly elaborate on the 
particular case where $B=0$, i.e. on the quantum algebra (\ref{nctheta}).
 This
algebra is realized by the operators (\ref{repthe}) which
act on $L^2 ({\bf R} ^2, dp_1 dp_2)$ as in equation (\ref{msr}).
Different representations of the quantum algebra,  
i.e. different
gauges for $\hat{\vec X}$ are related by a gauge
transformation described by a function $\tilde{\alpha} (\vec p \,
)$: $\tilde A_j \mapsto \tilde A_j^{\prime} = \tilde A_j +
{\partial \tilde{\alpha} \over \partial p_j}$. The induced
transformation of $\hx_i$ and $\hp_j$ is a unitary transformation:
$\hx _i ^{\prime} = U\hx _i U ^{-1}$ and $\hp _j ^{\prime} = U\hp
_j U ^{-1}$ with $U = \re ^{\ri \tilde{\alpha} (\vec{p} \, ) }$.
The wave function
$\tilde \psi (p_1, p_2)$
transforms according to
$\tilde{\psi} \mapsto
\tilde{\psi} ^{\prime} = U \tilde{\psi}
= \re^{\ri \tilde{\alpha} (\vec p \, )} \tilde{\psi} $
and the eigenvalue equation for
$H( \hat{\vec X} , \hat{\vec P} \, )$, i.e.
$H \tilde{\psi} = E \tilde{\psi}$, becomes
$H ^{\prime} \tilde{\psi}^{\prime} = E \tilde{\psi}^{\prime}$.
Thus, the spectrum of the
Hamiltonian $H( \hat{\vec X} , \hat{\vec P} \, )$
does not depend on the gauge chosen for $\hat{\vec X}$,
i.e. on the representation which is considered for the quantum algebra.

\section{NCQM systems: operatorial approach}

In this section, we discuss the standard {\em operatorial
approach} to NCQM. 
The {\em star product approach}
(``deformation quantization") will be treated in
section~\ref{starpro} while 
the {\em path integral approach} is to be
commented upon 
elsewhere~\cite{inprepa}.

The noncommutative Landau problem
(i.e. a particle in the noncommutative plane coupled to a {\em constant}
magnetic field which is perpendicular to this plane~\cite{nair})
can be treated
along the lines of its commutative counterpart.
Thus, two equivalent approaches can be considered:
either the Hamiltonian is expressed in terms of the canonical momentum
and a vector potential describing the magnetic field,
or the  Hamiltonian has the form of a free particle Hamiltonian,
but involving a momentum whose
components do not mutually commute, their
commutator being given by  the magnetic field strength.
The latter approach will be described in the first subsection.
We note that this formulation (which does not involve a gauge potential)
cannot be generalized to non-constant magnetic fields
since the Jacobi identities for the commutator algebra
are violated for $\theta \neq 0$ and $B$ depending on $\hat{\vec X}$
(see section 3.1.3).

In subsection~\ref{ahat},  we will discuss the approach relying on a vector
potential and we will see that the noncommutativity of the
configuration space variables then leads to expressions
which are characteristic for non-Abelian Yang-Mills (YM) theories.
A comparison between the different approaches to
magnetic fields in NCQM will be made in section 8.

As before and unless otherwise stated,
{\em the canonical operators
$\vec X$ and $\vec P$ are assumed to act on
$L^2 ({\br} ^2, dx_1dx_2)$ in the standard manner.}

\subsection{Commutation relations involving
a constant magnetic field}\label{ccrcmf}

For
a particle of unit mass and charge,
we consider the Hamiltonian
$H = \frac{1}{2} \, \hat{\vec P} ^{\; 2}$
which depends on momentum variables ${\hp}_j$
which do not commute with each other,
but rather satisfy the algebra  (\ref{thb}):
\[
[ \hx _1 , \hx _2 ] =  \ri \theta \Id
\, , \qquad
[ {\hp}_1 , {\hp}_2 ] = \ri  B \Id
\, , \qquad
[ {\hx}_i , {\hp}_j ] = \ri \delta_{ij} \Id
\, .
\]
The spectrum of $H$ may be determined by mimicking the reasoning
presented in equations (\ref{rewrite})-(\ref{speclandau}):
the calculation and the final result
\begin{equation}
\label{specnclandau}
E_n = |B| \, (n + {1 \over 2} )
\qquad {\rm with} \quad n \in \{ 0,1,2,\dots \}
\end{equation}
do not involve the parameter $\theta$ which means that the energy
levels are identical in standard quantum mechanics ($\theta =0$)
and in NCQM ($\theta \neq 0$) \footnote{Concerning this result, it
is worthwhile recalling from section 3.2.1 that the classical
equations of motion for $\theta =0$ and $\theta \neq 0$ are
also identical.}. 
This result reflects the
fact that the Hamiltonian does not depend on the coordinates
$\hx_1, \hx_2$ so that its spectrum is insensitive to the value of
the commutator $[ \hx_1, \hx_2 ]$. However, there are 
observable quantities which do not coincide in standard quantum
mechanics and in NCQM, e.g.~\cite{nair} the density of states per
unit area $\rho = \ds{1 \over  2\pi} \, \left| \ds{B \over 1 - B
\theta} \right|$. The latter diverges at the critical point $B = 1
/\theta$. A detailed study of the physical observables in the
noncommutative Landau system is presented in
reference~\cite{ricc}.

\subsection{Minimal coupling to a (possibly non-constant) 
magnetic field}\label{ahat}

We start~\cite{correa} from the algebra
describing the noncommutativity of
configuration space,
\begin{equation}
\label{nc12}
[\hx_1 , \hx_2 ] = \ri \theta \Id \, , \qquad
[\hp_1 , \hp_2 ] = 0  \, , \qquad
[\hx_i , \hp_j ] =  \ri \delta_{ij} \Id
\end{equation}
and from the Hamiltonian $H= {1 \over 2m} ( \hat{\vec P} - e \hat{\vec A} \, )^2$.
Here, $\hat{\vec A}$ is a vector field depending on the
non-canonical  operators $\hx_1, \hx_2$, i.e.
$\hat{\vec A} \equiv \vec A ( \hat{\vec X} \, )$,
with some ordering prescription for the variables $\hx_1$ and $\hx_2$.
The operator $\hx_i$ is to be thought of
as a function of the canonical  operators
$X_j$ and $P_k$
(see equation (\ref{repthe}) for a general expression),
while $\hat{\vec P} = {\vec P} = -\ri  \vec{\nabla}$.
From relations (\ref{nc12}), it follows that the components of the kinematical
momentum $ \hat{\vec \Pi} = \hat{\vec P} - e \hat{\vec A}$ satisfy
\begin{equation}
\label{ncpi}
[\hat \Pi _1 , \hat \Pi _2 ] = \ri e \hat F _{12} \Id \, , \qquad
{\rm with} \ \;
 \hat F _{12} \equiv \partial _1 \hat A _2 - \partial _2 \hat A _1
- \ri e \, [ \hat A _1 ,\hat A _2 ]
\end{equation}
where $\partial_j \equiv \partial \, / \partial x_j$.
We note that $\hat F _{12}$ has the same expression as the
field strength in non-Abelian YM theories.
Since $\hx_i$ depends on $\vec P$ (according to (\ref{repthe})),
one has $[\hx_i,  \hat \Pi _j ] = \ri \delta _{ij} \Id$ {\em plus}
additional terms (specified in equation (\ref{ncgal})
below within the star product approach).

Since the vector potential depends on the noncommuting variables
$\hx _i$, 
it is natural to assume that the gauge transformations
also depend on $\hat{\vec X}$. Thus, the wave function
$\psi (\vec x \, )$ transforms as
$\psi \mapsto \psi ^{\prime} = \hat U \psi$ with
$\hat U = \re ^{\ri \lambda (\hat{\vec X} \, )}$.
By requiring the covariant derivative
$\hat \Pi _i \psi$ to transform in the same manner as $\psi$, one finds
\begin{eqnarray}
\hat A_i ^{\prime} \!\!\! &=&  \!\!\!
 \hat U \hat A_i \hat U^{-1}
+ {\ri \over e} \, \hat U \partial_i \hat U^{-1}
\nonumber
\\
\hat F_{ij} ^{\, \prime}  \!\!\! &=&  \!\!\!
 \hat U \hat F_{ij} \hat U^{-1}
\, .
\label{trafofij}
\end{eqnarray}
Henceforth, the consideration of noncommuting coordinates
in configuration space yields transformation laws
 for the gauge fields
that are characteristic for a non-Abelian gauge theory.
If the noncommutative field strength $\hat F_{12}$
is constant 
(i.e. independent of $\hat{\vec X}$), 
it is gauge invariant by virtue of equation
(\ref{trafofij}).

More specifically, let us now consider the symmetric gauge
for $\hat{\vec A} $,
\begin{equation}
\label{bbar1}
\hat{\vec A} = \left( - {\bar B \over 2} \,  \hx_2 ,
{\bar B \over 2} \,  \hx_1 \right)
 \, ,
\end{equation}
where $\bar B$ is a constant (which might depend on $\theta$).
Substitution of (\ref{bbar1}) into (\ref{ncpi}) yields
\begin{equation}
 \label{ncfst}
\hat F _{12} = \bar \Lambda \bar B \, , \qquad {\rm with} \ \;
\bar \Lambda = 1 + {e \over 4} \theta \bar B
\, .
\end{equation}
If $\bar B$ depends on $\theta$ and on a $\theta$-independent constant
$B$ according to
\begin{eqnarray}
\label{bbar}
\bar{B} = \bar{B} (B; \theta)
\!\!\! & \equiv & \!\!\!
 {2 \over e \theta}
\left( \sqrt{1 + e \theta B}  - 1 \right)
\\
\!\!\! & = & \!\!\!
B \, ( 1 + {e \over 4} \, \theta B) + {\cal O} (\theta^2)
\, ,
\nonumber
\end{eqnarray}
then relation (\ref{ncfst}) implies
that the noncommutative field strength $\hat F _{12} $ is a
$\theta$-independent constant: $\hat F _{12} = B$.
In this case, we have the same algebraic setting as in
subsection~\ref{ccrcmf}:
the Hamiltonian reads as
$H = {1 \over 2m} \hat{\vec \Pi}^2$ and depends
on the non-canonical momentum components $\hat{\Pi}_j$
which satisfy
$[\hat \Pi _1 , \hat \Pi _2 ] = \ri e B \Id$
(where $B$ is a  $\theta$-independent constant).
Thus, the spectrum of $H$ is the same as in ordinary quantum mechanics:
$E_n = {| e B | \over m} \, (n+{1 \over 2})$ with $n =0,1,\dots$.

The same result can be obtained by considering the Landau gauge
$\hat{\vec A} = \left( 0, B \hx_1
\right)$ which yields the same constant
field strength: $\hat F_{12} = B$.
By contrast to the star product formalism
described in section~\ref{magfield} below,
the approach described in this subsection
does not fix afore-hand a
specific gauge for $\hat{\vec X}$ 
(i.e. a specific representation of $(\hat{\vec X} , \hat{\vec P}\, )$
in terms of $( {\vec X} , {\vec P}\, )$)
and it works in a simple way
for any choice of gauge potential $\hat{\vec A}$
describing a constant magnetic field strength $\hat F_{12}$.
The commutation relations for $\hat{\vec X}$ and
$\hat{\vec \Pi}$ have the form (\ref{ncgal})
and they allow for a treatment of variable magnetic fields.

\subsection{``Exotic'' approach}\label{exotic}

We note that there exists another description of the
NC Landau system, namely
the so-called {\em ``exotic'' model}~\cite{horvathy,duval,duvalhor}.
This approach is based on the Poisson algebra (\ref{exob}), 
i.e. it amounts to modifying the commutators   (\ref{thb})
by a factor $1/\kappa$.
Applications of the present model 
to the fractional quantum Hall effect
and to the vortex dynamics
in a thin superfluid $^4{\rm He}$ film
are discussed in reference~\cite{horvathy}.

\section{NCQM systems: star product approaches}\label{starpro}

Consider
the CCR's (\ref{heisen}) of standard {\bf quantum mechanics}
in two dimensions: 
\[
[ X _1 , X _2 ] = 0=
[ P_1 , P_2 ]
\, , \qquad
[ X_i , P_j ] = \ri \delta_{ij} \mathds{1}
\qquad {\rm for} \ \; i,j \in \{ 1, 2 \}
\, .
\]
In the operatorial approach, one works with functions of the
noncommuting operators $X_i, P_j$ and with the ordinary
product of such functions.
Alternatively, one can consider functions
$f(\vec x , \vec p \, )$
depending on the ordinary
commuting coordinates $x _i, p_j $ and multiply
these functions by a noncommutative product,
namely the so-called star product.
This formalism,
initiated by Weyl and Wigner~\cite{weyl},
developped by Groenewold and Moyal~\cite{groen}
and further generalized in the
sequel~\cite{bayen, kontse,zachos}
allows for a phase space description of quantum mechanics.
This autonomous approach to quantum mechanics
is also referred to as
{\em deformation quantization} - see~\cite{zachos,walton,hirsh}
for a nice summary and overview.
Here, we only note that the original formalism
of Groenewold and Moyal
assumes Weyl ordering of the noncommuting variables.

The procedure of deformation quantization
can be generalized in different manners
to {\bf NCQM} as described (in its simplest form) by
the algebra
\begin{equation}
\label{ncftcopy}
[ \hx _1 , \hx _2 ] = \ri \theta \Id
\, , \qquad
[ {\hp}_1 , {\hp}_2 ] = 0
\, , \qquad
[ {\hx}_i , {\hp}_j ] = \ri \delta_{ij} \Id 
\qquad {\rm for} \ \; i,j \in \{ 1, 2 \}
\, .
\end{equation}
The obvious generalization consists of
defining a star product for
phase space functions $f(\vec x , \vec p \, )$
which implements all of the commutation relations
(\ref{ncftcopy}).
In this formulation, a quantum state is also a
function on phase space, namely the so-called Wigner function.
Such a phase space formulation of
NCQM has been considered by some authors, 
e.g.~\cite{hatzi, ncqmphase}.

Another approach~\cite{mezin}, to be treated in detail in this section,
proceeds along the lines of the
star product formulation of field theory
on noncommutative space~\cite{filk}:
here one only introduces a
{\bf star product in configuration space}
in order to implement the noncommutativity of the
$x$-coordinates, i.e. the first of relations
(\ref{ncftcopy}).
By definition,
the {\bf Groenewold-Moyal star product}
of two smooth functions $V$ and $\psi$
depending on $\vec x$
is a series in $\theta$ given by
\begin{equation}
\label{starp}
V \star \psi = V \psi
+ {1 \over 1!}  \, {\ri \over 2} \, \theta ^{ij}  \, \partial _i V \, \partial _j \psi
+ {1 \over 2!}  \, \left( {\ri \over 2} \right) ^2  \,\theta ^{ij} \theta ^{kl} \,
\partial _i \partial _k V \, \partial _j \partial _l \psi
+ \dots
\, ,
\end{equation}
which implies that
\[
[x_1 \stackrel {\star}{,} x_2 ] \equiv x_1 \star x_2 - x_2 \star x_1
= \ri \theta
\, .
\]
For this formulation of NCQM,
quantum states are described as usual by 
wave functions $\psi(\vec x \,)$ on configuration space
and
the momentum variables are also assumed to act as usual
on these functions, i.e. $P_j = -\ri \, \partial / \partial x_j$
(as in the representation (\ref{repthe})
 of the commutation relations (\ref{ncftcopy})).

Before considering a particle subject to a magnetic field, 
we investigate the simpler case of a particle 
in a scalar potential.

\subsection{Scalar potentials}\label{stap}

In the standard formulation of
ordinary quantum mechanics,
the potential energy $V$
acts on the wave function $\psi \in L^2 ({\bf R} ^2, dx_1 dx_2)$
as an operator of multiplication: $\psi \mapsto V \cdot \psi$.
Accordingly, in the star product formulation of NCQM,
the potential energy $V$
acts on the wave function $\psi$
by the star product (\ref{starp}).
With the help of the Fourier transform, one
can check~\cite{mezin, bigatti} that the expression (\ref{starp})
can be rewritten as
\begin{equation}
\label{starop}
\left( V \star \psi \right) ( \vec x \, )
= V ( \hat{\vec X}  \, ) \, \psi ( \vec x \, )
\qquad {\rm with} \ \, \hat X_i \equiv X_i - {1 \over 2} \theta^{ij} P_j
\, .
\end{equation}
Here, the quantity $V ( \hat{\vec X} \, )$
is to be considered as an operator acting on the wave function
$\psi  \in L^2 ({\bf R} ^2, dx_1 dx_2)$.
In particular, we have
\begin{equation}
\label{xstar}
\left( x_i   \star \psi \right) ( \vec x \, ) =
\hat X _i \, \psi ( \vec x \, )
\, .
\end{equation}
Thus, we recover the representation 
\[
\hat X _i = X _i - {1 \over 2} \, \theta^{ij} P_j
\, ,
\qquad \hat P_j = P_j
\, , 
\]
of the algebra (\ref{ncftcopy}), 
i.e. the ``symmetric gauge'' for $\hat{\vec X}$.

Obviously, equation (\ref{starop}) corresponds to a particular
ordering prescription: the right-hand-side
is defined in terms of the left-hand-side which is unambiguous.
One can verify that this prescription is indeed the one of Weyl:
by spelling out $(x_i x_j) \star \psi$
with the help of (\ref{starp}), one finds
\begin{equation}
\label{weylorde}
\left( ( x_i x_j) \star \psi \right)  ( \vec x \, ) =
{1 \over 2} \, ( \hx_i \hx_j + \hx_j \hx_i )
 \, \psi ( \vec x \, )
\, .
\end{equation}

{\em In summary:} For NCQM based on the algebra (\ref{ncftcopy}),
there are two equivalent approaches
(at least for sufficiently regular potential functions):
first, the operatorial approach in which
the Hamiltonian is expressed in terms
of the operators $\hp_j \equiv P_j \equiv -\ri \partial_j$
and in terms of the non-canonical
operators $\hat X _i$
and, second, the star product approach in which
the Hamiltonian is written in terms
of $P_j \equiv -\ri \partial_j$ and real
coordinates $x_i$ acting on the wave function
by the Groenewold-Moyal star product (\ref{xstar}).
The latter formulation relies on the {\em Weyl ordering}
of the variables $\hat X _i$ 
and corresponds to the choice of the
{\em symmetric gauge} for $\hat{\vec X}$, see equations
(\ref{starop})-(\ref{weylorde}).
(For a different ordering prescription, namely normal ordering
(hence a different star product),
we refer to \cite{ouvry}.)

For the Hamiltonian
$H= {1 \over 2m} \, \vec p ^{\; 2} + V(\vec x \, )$
acting on wave functions by the star product, 
the probabilistic interpretation  proceeds along the usual lines,
all products being replaced by star products.
Indeed, the time derivative of the probability density
$\rho \equiv \bar{\psi} \star \psi$
can be determined by using the Schr\"odinger equation
\[
H\star \psi = \ri \hbar \, \partial _t \psi
\, , \qquad {\rm with} \ \; H= {1 \over 2m} \, {\vec p}^{\; 2} + V(\vec x \, )
\]
and its complex conjugate (taking into account the fact that
$\overline{V \star \psi } = \bar{\psi} \star V$ for the real-valued
potential $V$). Thus, one obtains the continuity equation
\[
 \partial _t \rho + {\rm div} \, \vec{\jmath} = 0
\, , \quad {\rm with} \ \,
\left\{
\begin{array}{l}
\rho = \bar{\psi} \star \psi \\
\vec{\jmath} = {\hbar \over 2m \ri} \, [ \, \bar{\psi} \star \vec{\nabla} \psi
- (\vec{\nabla}  \bar{\psi}) \star \psi \, ] \, ,
\end{array}
\right.
\]
which implies that the total probability
$\int_{{\bf R} ^2} d^2x \,  \bar{\psi} \star \psi$ is a
conserved quantity.

\subsection{Magnetic fields}\label{magfield}

Let us now consider
a particle moving in the noncommutative plane
subject to a (possibly non-constant) magnetic field
which is perpendicular to this plane.
The present section describes two  approaches
to this problem which are both based on the star product
in configuration space and on the introduction of
a vector potential (i.e. a $U(1)$ gauge field).
The first approach solely relies on the star product~\cite{mezin, dayi}
while the second one supplements the star product with
the so-called Seiberg-Witten map~\cite{chakra,kokado,harikumar, scholtz}.
The results  obtained for the particular case of a constant magnetic field
(i.e. the Landau system) will be compared with those
obtained for this system in section 6.

\subsubsection{Gauge potentials}\label{gaugepot}

We use the following notation.
The fields of NCQM (which
depend on the coordinates $\vec x = (x_1, x_2)$
and which are multiplied by each
other by means of the star product) are  denoted by a
`check'\footnote{We do not use the standard notation~\cite{sw},
i.e. a `hat', in order to avoid any confusion with the
hatted operators $\hx_i$ and $\hp_j$ considered
in the earlier sections.}
e.g. $\check{\psi}, \check A_i, \check \lambda,\dots$.

A local $U(1)$ gauge transformation of the wave function $\check \psi$
(describing a particle of charge $e$)
is given by
\begin{equation}
\label{ncgt1}
\check \psi \longmapsto \check \psi^{\prime} = \check U_{\check \lambda}
\star \check \psi
\, , \qquad {\rm with} \ \;
\check U_{\check \lambda} = \re _{\star} ^{\ri e \check \lambda}
\, ,
\end{equation}
where all products of $\check \lambda$ in the exponential
$\check U_{\check \lambda}$ are star products.
The inverse of $\check U_{\check \lambda}$
is determined by the relation
$\check U_{\check \lambda} \star \check U_{\check \lambda}^{-1} =1$.
The covariant derivative of $\check \psi$, as defined by
$\check D_i \check \psi = \partial_i \check \psi
- \ri e \check A_i \star \check \psi$,
transforms in the same way as $\check \psi$ if the vector
potential $\check A_i$ transforms according to
\begin{equation}
\label{ncgt2}
\check A_i  \longmapsto \check A_i^{\prime} =
 \check U_{\check \lambda} \star \check A_i \star
\check U_{\check \lambda}^{-1}
+ {\ri \over e} \,  \check U_{\check \lambda} \star
(\partial_i   \check U_{\check \lambda}^{-1})
\, .
\end{equation}
The commutator of the covariant derivatives
determines the magnetic field strength:
\begin{equation}
{[ \check D_i  \stackrel{\star}{,} \check D_j ]}
\check \psi
= -\ri e \check F_{ij} \star \check \psi
\, , \qquad
{\rm with} \ \;
\check F_{ij} = \partial_i  \check A _j - \partial_j  \check A_i
- \ri e \, [  \check A_i \stackrel{\star}{,} \check A_j ]
\, .
\label{ncfs}
\end{equation}
The latter transforms according to
\begin{equation}
\label{ncgt3}
\check  F_{ij}  \longmapsto \check F_{ij}^{\, \prime} =
 \check U_{\check \lambda} \star \check F_{ij}
\star   \check U_{\check \lambda}^{-1}
\, .
\end{equation}
Thus, for the Abelian gauge group  $U(1)$,
the noncommutativity of the star product
induces expressions that are characteristic
for non-Abelian Yang-Mills theories~\cite{mezin, nekra,bigri}.
For the two-dimensional configuration space
that we consider here for simplicity, the only non-vanishing
component of the tensor $\check F_{ij}$ is
$\check F_{12} \equiv B_*$.
By contrast to the field strength
$F_{ij} = \partial_i A_j -  \partial_j A_i$
of an ordinary $U(1)$ gauge theory, the
field strength $\check F_{ij}$ is not gauge invariant,
but transforms covariantly as in equation (\ref{ncgt3}).
Instead of the fundamental representation for the matter field
(that we consider here and for which the covariant derivative
reads as $\check D_i \check \psi = \partial_i \check \psi
- \ri e \check A_i \star \check \psi$),
one can also consider the anti-fundamental or the adjoint
representation (for which we have, respectively,
$\check D_i \check \psi \equiv \partial_i \check \psi
+ \ri e \check \psi \star \check A_i $ and
 $\check D_i \check \psi \equiv \partial_i \check \psi
- \ri e [ \check A_i \stackrel{\star}{,}  \check \psi ]$).

The {\em Schr\"odinger equation}
$\check H \star \check \psi = \ri \partial_t \check \psi$
with
\begin{equation}
\label{hsvp}
\check H = {1 \over 2m} \, (\vec P - e \check {\vec A} \, )^2_{\star}
\, \equiv \,
{1 \over 2m} \sum_i (P_i - e \check A_i) \star (P_i - e \check A_i)
\end{equation}
is invariant under the gauge transformations (\ref{ncgt1}),(\ref{ncgt2})
with a gauge parameter $\check \lambda$ which does not depend on time.
Let us now consider the case where the field strength $\check F_{ij}$
is {\em constant}.
This case is quite particular since it follows from (\ref{ncgt3})
that such a field $\check F_{ij}$
is gauge invariant.

Let us choose~\cite{mezin,dayi} the  symmetric gauge
\begin{equation}
\label{vp}
\check{\vec A} (\vec x \, ) =
\left( -{\bar B \over 2} x_2 , {\bar B \over 2} x_1 \right)
\, ,
\end{equation}
where the real constant $\bar B$ parametrizes the magnetic
field.
From (\ref{ncfs}), we infer that the associated field strength
is given by
\begin{equation}
\label{lamth}
\check F_{12} =
\bar \Lambda \bar B
\, , \qquad {\rm with} \ \;
\bar \Lambda = 1 + {e \over 4} \, \theta \bar B
\, .
\end{equation}
Henceforth, we presently have expressions of the same form
as in the operatorial approach -- compare
(\ref{vp}) with (\ref{bbar1})
and (\ref{lamth}) with (\ref{ncfst}).

The star products
can be disentangled by virtue of relation (\ref{xstar}):
\begin{equation}
\label{dise}
(P_i - e \check A_i ) \star \check \psi = \bar \Lambda
\,
\left( P_i - {e \over \bar \Lambda }\,  \check A_i (\vec x \, ) \right) \check \psi
\, .
\end{equation}
By applying this relation once more, one finds~\cite{mezin}
that $\check H \star \check \psi$
(with $\check A_i$ given by (\ref{vp})) reads as
\begin{equation}
\label{diseh}
\check H \star \check \psi = H_* \check \psi
\, ,
\end{equation}
with
\[
H_* = {1 \over 2 m_*} \,
\left( \vec P - e_* \check{\vec A} (\vec x \, ) \, \right)^2
, \quad {\rm where} \ \;
m_* = \bar \Lambda^{-2} m
\, , \ \;
e_* = \bar \Lambda^{-1} e
\, .
\]
Here, we put a lower index $*$
on the quantities which
have been modified due to the consideration of star products
(i.e. the quantities $H_*, m_*$ and $e_*$).

The Hamiltonian $H_*$ has the same form as the corresponding Hamiltonian
in ordinary quantum mechanics
(the only difference being 
a redefined mass and charge which depend on $\theta$).
As a consequence of this fact and of the relation
$\partial _1 \check A_2 - \partial _2 \check A_1 =\bar B$,
the spectrum of $H_*$
is given, in analogy to expression (\ref{speclandau}),  by
\begin{eqnarray}
E_n \!\!\! &=&  \!\!\!
\hbar \, {|e_* \bar B| \over m_* c} \, ( n+ {1 \over 2} )
\nonumber \\
\!\!\! &=&  \!\!\!
\hbar \, {|e \check F_{12} | \over m c} \, ( n+ {1 \over 2} )
 \qquad
 {\rm with} \ \;  n \in \{ 0, 1, \dots \}
\, .
\label{landstar}
\end{eqnarray}
These energy levels are the same as in ordinary quantum mechanics, but
with the magnetic field
$\bar B = \partial _1 \check A_2 - \partial _2 \check A_1$ replaced by
the noncommutative field strength
$\check F_{12}$.
The latter field depends on $\theta$ according to (\ref{lamth})
except for the case where  $\bar B$ (and thereby $\check{\vec A}$)
depends on $\theta$ in such a way that
$\check F_{12}$ is a $\theta$-independent constant
$B$: if relation (\ref{bbar}) holds, i.e. if 
\begin{eqnarray}
\label{bbarcopy}
\bar{B} = \bar{B} (B; \theta)
\!\!\! & \equiv & \!\!\!
 {2 \over e \theta}
\left( \sqrt{1 + e \theta B}  - 1 \right)
\\
\!\!\! & = & \!\!\!
B \, ( 1 + {e \over 4} \, \theta B) + {\cal O} (\theta^2)
\, ,
\nonumber
\end{eqnarray}
then equation (\ref{ncfs}) yields $\check F_{12}= B$.
The symmetric gauge (\ref{vp})
with $\bar B$ given by (\ref{bbarcopy})
is gauge equivalent to the Landau gauge  
$\check{\vec A} = (0, B x_1)$.

{\em In summary:}
There are two different view-points for the noncommutative Landau system
described in terms of a vector potential and the star product.
The first view-point is that the variable $\bar B$
in the vector potential (\ref{vp}) is a $\theta$-independent constant:
this leads to a $\theta$-dependent energy spectrum
($E_n \propto | \bar \Lambda \bar B | $) which differs
from the one that we obtained
by the approach of section~\ref{ccrcmf}
($E_n \propto | \bar B |$)
where we considered a free Hamiltonian and
a commutator of momenta determined by the constant magnetic field:
$[ \hp _1, \hp _2] \equiv \ri \bar B \Id$.
The second view-point consists of regarding the noncommutative
field strength $\check F_{12}$ 
as the central variable, thereby considering
this field as a $\theta$-independent constant:
$\check F_{12}= B$.
In this case, the vector potential (\ref{vp}) depends on
$\theta$ according to (\ref{bbarcopy}).
This implies that the energy spectra
obtained, respectively,
by the approach of this section and the one of section~\ref{ccrcmf}
(with  $[ \hp _1, \hp _2] \equiv \ri B \Id$) coincide with each other:
$E_n \propto | B |$. This spectrum also coincides
with the one of ordinary quantum mechanics.

\medskip

\noindent
{\bf Remarks:}

\smallskip

{\bf (i)}
As was already pointed out in reference~\cite{mezin},
the argumentation (\ref{dise}),(\ref{diseh})
 relies on the symmetric gauge (\ref{vp})
for $\check{\vec A}$ (e.g. equation (\ref{dise}) does not hold
in a Landau-type gauge).
This result is related to the fact that 
the star product amounts
to choosing the symmetric gauge for $\hat{\vec X}$ 
-- see equation (\ref{xstar}).
Henceforth, the spectrum for the minimally coupled Hamiltonian
can only be determined by simple algebraic arguments if the
symmetric gauge is chosen as well 
for the vector potential $\check{\vec A}$.
Other gauge choices require more complicated computations. 

\smallskip

{\bf (ii)}
The gauge invariance of the $\star$-eigenvalue equation
$\check H \star \check \psi = E_n \check \psi$
is ensured by construction.
The ordinary eigenvalue equation
$\check H_{\ast} \check \psi = E_n \check \psi$
is invariant under ordinary gauge transformations
of $\check{\vec A}$ and $\check \psi$,   
the relationship between the different gauge transformations
being the subject of the Seiberg-Witten map
discussed in the next subsection. 

\smallskip

{\bf (iii)}
If a scalar potential $\check V$ is added to the Hamiltonian
(\ref{hsvp}),
the function $\check V \star \check \psi$
transforms in the same manner as $\check \psi$
under a local $U(1)$ gauge transformation $\check U _{\check \lambda}$,
if the potential $\check V$
transforms according to
$\check V \mapsto \check V ^{\prime} =
 \check U_{\check \lambda}
\star \check V \star  \check U_{\check \lambda}^{-1}$ \cite{scholtz}.
The example of a constant electric field
$\vec E = - \vec \nabla V$ is treated in reference~\cite{dayi}
and applied to the Hall effect and the Aharonov-Bohm effect.

\smallskip

{\bf (iv)}
If we introduce the components of the kinematical momentum,
$\check \Pi _k \equiv -\ri \, \check D_k
= - \ri \partial _k -e \check A_k$,
we obtain the {\em $\star$-commutation relations}
\begin{eqnarray}
{[x_1 \stackrel{\star}{,} x_2 ]}
\!\!\! &=&  \!\!\!
 \ri \theta
\nonumber
\\
{[\check \Pi _1 \stackrel{\star}{,} \check \Pi _2 ]}
\!\!\! &=&  \!\!\!
\ri e \, \check F_{12} \, ,
\qquad \quad {\rm with}
\ \,
\check F_{12} = \partial_1 \check A_2 - \partial_2 \check A_{1}
-\ri e \, [  \check A_{1}  \stackrel{\star}{,} \check A_2 ]
\nonumber
\\
{[x_i \stackrel{\star}{,} \check \Pi _j ]}
\!\!\! &=&  \!\!\!
 \ri \, \delta_{ij} - e \,
{[x_i \stackrel{\star}{,} \check A _j ]}
\, .
\label{ncgal}
\end{eqnarray}
By contrast to the commutation relations (\ref{thb})
involving a constant magnetic field $B$, i.e.
\begin{equation}
\label{thbbis}
[ \hx _1 , \hx _2 ] =  \ri \theta \Id
\, , \qquad
[ {\hp}_1 , {\hp}_2 ] = \ri  B \Id
\, , \qquad
[ {\hx}_i , {\hp}_j ] = \ri \delta_{ij} \Id
\, ,
\end{equation}
we may presently have a
{\em non-constant magnetic field,} but the algebra (\ref{ncgal})
contains some extra terms on the right-hand side
which are due to the fact that $x_1$ and $x_2$ do not $\star$-commute.
These additional terms ensure the validity of the Jacobi
identities for the algebra (\ref{ncgal})
for any vector potential $\check{\vec A}$.
In particular, the Jacobi
identities hold for a vector potential
describing a variable magnetic field
by contrast to the algebra (\ref{thbbis})
for which these identities
do not hold if $B$
depends on $\hat{\vec X} \, $. Yet,
if the field strength $\check F_{12}$ is not constant,
it is not gauge invariant which renders
its physical interpretation
as a magnetic field somewhat unclear.
As we will see in the next subsection,
this issue can be settled by considering the Seiberg-Witten map
which relates the
noncommutative gauge fields and parameters
to their   commutative counterparts
in a specific way.

\subsubsection{Gauge potentials and Seiberg-Witten map}\label{swmap}

In the previous subsection, we saw that the consideration of star
products in the Schr\"odinger equation coupled to a $U(1)$ gauge field
leads to the emergence of
a non-Abelian gauge structure in NCQM which is not present in 
ordinary quantum mechanics.
The situation is completely
analogous in relativistic field theory
when passing from an  ordinary  $U(1)$ gauge theory
to a noncommutative $U(1)$ gauge theory.
In the latter context, Seiberg and Witten~\cite{sw} have
introduced a map between both theories
which relates noncommutative gauge fields and parameters
to their commutative
counterparts in such a way that
gauge equivalent field configurations are mapped into each other.
By virtue of this map, both theories can be considered
as {\em equivalent}
and the action functional or Hamiltonian of the noncommutative  $U(1)$ theory
can be expressed in terms of ordinary $U(1)$ gauge fields
and the noncommutativity parameters $\theta_{ij}$
(which may be viewed as a constant ``background field'').

In this section, we will discuss this so-called Seiberg-Witten map
within the context of quantum
mechanics~\cite{chakra,kokado,harikumar, scholtz}.
We use the same notation as in the previous subsection,
i.e. the fields of NCQM
(which depend on the coordinates $\vec x = (x_1, x_2)$
and which are multiplied with each
other by means of the star product) are  denoted by a
`check'. We also spell out the coupling constant $e$
which is usually absorbed into the fields.

By definition, the {\em Seiberg-Witten map} is a mapping
$A_i \mapsto \check{A}_i (A)$
and $\lambda  \mapsto \check \lambda (\lambda, A)$
(the `checked' fields being power series in
the noncommutativity parameters $\theta_{ij} = \theta \varepsilon_{ij}$)
which satisfies
\begin{equation}
\label{swm}
\check A_i (A) + \check{\delta}_{\check{\lambda} } \check A_i (A)
= \check A_i (A+ \delta_{\lambda} A)
\, .
\end{equation}
Here, $\check{\delta}_{\check{\lambda} } \check A_i =
\check D_i \check{\lambda} \equiv \partial _i \check{\lambda}
- \ri  e \, [  \check A_i \stackrel{\star}{,} \check \lambda \, ]$ and
$\delta_{\lambda} A_i = \partial _i \lambda$
describe infinitesimal $U(1)$ gauge transformations
in NCQM and in ordinary quantum mechanics, respectively.
Equation (\ref{swm}) is solved~\cite{sw} by
\begin{eqnarray}
A_i \!\!\! & & \!\!\! \longmapsto \ \check{A}_i (A; \theta) = A_i
- {e \over 2} \, \theta^{kl} A_k (\partial_l A_i +F_{li})
+{\cal O} (\theta^2)
\\
\lambda  \!\!\! & & \!\!\! \longmapsto \ \check{\lambda}
(\lambda, A; \theta) = \lambda
- {e \over 2} \, \theta^{ij} A_i  \partial_j \lambda
+{\cal O} (\theta^2)
\, ,
\nonumber
\end{eqnarray}
and we have analogous maps~\cite{bichl} for the matter field $\psi$ and
the field strength of $A$:
\begin{eqnarray}
\psi \!\!\!&&\!\!\! \longmapsto \ \check{\psi} (\psi, A; \theta) = \psi
- {e \over 2} \, \theta^{ij} A_i  \partial_j \psi
+{\cal O} (\theta^2)
\label{mattf}
\\
F_{ij} \!\!\!&&\!\!\! \longmapsto \ \check{F}_{ij} (A; \theta) = F_{ij}
+  e \theta^{kl} F_{ik} F_{jl}
+{\cal O} (\theta^2)
\nonumber
\, .
\end{eqnarray}
Here, the expression for $\check{F}_{ij}$
follows from the defining relation (\ref{ncfs}).
The fact that $\check \lambda$ not only depends on $\lambda$,
but also on $A$ implies there is no well-defined
mapping between the gauge group
$\{ U_{\lambda} = \re ^{\ri e \lambda} \}$
of ordinary gauge theory and the  gauge group
$\{ \check U_{\check \lambda} =
\re ^{\ri e \check{\lambda} } _{\star} \}$
of noncommutative  gauge theory~\cite{sw}.

For a {\em constant} magnetic field $\Bc \equiv F_{12}$,
one has the following exact (i.e. valid
to all orders in $\theta$) result  for the
noncommutative field strength $\check F _{12} \equiv \check B$
which is determined by the Seiberg-Witten map~\cite{sw}:
\begin{equation}
\label{ncfs2}
\check{B} = {\Bc \over 1 - e \theta \Bc}
\ .
\end{equation}

\bigskip

Just as the ordinary magnetic field $\Bc$ can be described
by a symmetric gauge field configuration, $A_i =  -{1 \over 2} \, \Bc
\varepsilon_{ij} x_j$ (with $i \in \{1,2\}$),
the Seiberg-Witten field strength $\check B \equiv \check F _{12}$
can be described by a symmetric field configuration~\cite{scholtz}:
\begin{eqnarray}
\label{ncsg}
\check{A}_i (A; \theta) = -{1 \over 2} \, \bar{B} (\Bc ; \theta) \,
\varepsilon_{ij} x_j
\, , \qquad {\rm with} \quad
 \bar{B} (\Bc ; \theta)
\!\!\! & = & \!\!\!
 {2 \over e \theta}
\left( {1 \over \sqrt{1 - e \theta \Bc } } - 1 \right)
\\
\!\!\! & = & \!\!\!
 \Bc ( 1 + {3 \over 4} \, e \theta \Bc ) + {\cal O} (\theta^2)
\, .
\nonumber
\end{eqnarray}
Indeed, substitution of this expression into $\check F _{ij}$, as defined
by relation (\ref{ncfs}), yields the
result (\ref{ncfs2}).
The relationship with the expressions in the previous subsection
can easily be worked out\footnote{
In fact,
in the present section we wrote
${\cal B} \equiv F_{12}$ rather than
$B \equiv F_{12}$ since the variable $B$ was used
in section~\ref{gaugepot}
with a different meaning. As in
subsection~\ref{gaugepot}, we presently have
$\check F_{12} = \bar{\Lambda} \bar B$ with
$\bar{\Lambda} = 1 + {e \over 4} \, \theta \bar B$. If
${\cal B} =   {B} \, (1 + e \theta B)^{-1}$, we recover the
 expressions of subsection~\ref{gaugepot}:
$\bar B$ as a function of $B$ is then given by equation (\ref{bbarcopy})
and $\check F_{12} = B$.}.

To discuss the Landau problem, we again proceed as in the previous
subsection, see equations (\ref{hsvp})-(\ref{landstar}).
Thus, we consider the Hamiltonian
\begin{equation}
\label{hsvp1}
\check H = {1 \over 2 \check m} \, (\vec P - e \check {\vec A} \, )^2_{\star}
\, \equiv \,
{1 \over 2 \check m} \sum_i (P_i - e \check A_i) \star (P_i - e \check A_i)
\, ,
\end{equation}
acting on the Seiberg-Witten field $\check \psi$.
In expression (\ref{hsvp1}), the Seiberg-Witten gauge field  $\check A_i$
describing a constant magnetic field
is assumed to be given by the symmetric
gauge (\ref{ncsg}).
The mass parameter has been denoted by $\check m$.
Very much like the Seiberg-Witten fields,
it may be viewed
as a function of $\theta$ satisfying $\check m (\theta =0) = m$.

By using relation (\ref{xstar}) to disentangle the star products,
we get
\[
\check H \star \check \psi = H_{{\rm SW}} \, \check \psi
\, ,
\]
with
\begin{equation}
\label{hamsw}
H_{{\rm SW}} = {1 \over 2 m_{{\rm SW}}} \,
\left( \vec P - e_{{\rm SW}} \check{\vec A} (\vec x \, )  \right)^2
, \quad {\rm where} \ \;
m_{{\rm SW}} = \bar{\Lambda}^{-2} \check m
\, , \ \;
e_{{\rm SW}} = \bar{\Lambda}^{-1} e
\, .
\end{equation}
The Hamiltonian $H_{{\rm SW}}$
(which depends on the constant ``background field'' $\theta$)
has the same form as
the corresponding Hamiltonian in ordinary quantum mechanics.
As a consequence of this fact and of the relation
$\partial _1 \check A_2 - \partial _2 \check A_1 =\bar B$, the spectrum of $H_{{\rm SW}}$
is given by
\[
E_n \, = \, \hbar \, {|e_{{\rm SW}} \bar B| \over m_{{\rm SW}} c}
\, ( n+ {1 \over 2} )
\, = \,
\hbar \, {|e \check B| \over \check m c} \, ( n+ {1 \over 2} )
 \qquad
 {\rm with} \ \;  n \in \{ 0, 1, \dots \}
\, .
\]
If we assume that
\begin{equation}
\label{massren}
\check m = {m \over 1 - e \theta \Bc}
\, ,
\end{equation}
then it follows from (\ref{ncfs2}) that
the energy levels read as
\begin{equation}
\label{specsw}
E_n = \hbar \, {|e \Bc | \over m c} \, ( n+ {1 \over 2} )
 \qquad
 {\rm with} \ \;  n \in \{ 0, 1, \dots \}
\, .
\end{equation}
This result coincides with the one of ordinary quantum mechanics
(for a particle of mass $m$ and charge $e$ coupled to the constant
magnetic field $\Bc = \partial _1 A_2 - \partial _2 A_1$).
In this sense, both theories are physically equivalent.
Note that the singularity $\Bc = (e \theta)^{-1}$
of the parameters $\bar B$ and $\check m$
does not manifest itself in the spectrum (\ref{specsw}).

 A somewhat different treatment of the Landau problem
within the Seiberg-Witten framework
(leading to the same result for the spectrum)
is presented in reference~\cite{scholtz}.
In the latter treatment, relation (\ref{massren})
follows from the requirement that the Hamiltonians
$H_{\rm SW} (\theta)$ and $H_{\rm SW} (0)$
are related by a unitary transformation
which ensures that the physics remains invariant
under a change in $\theta$.
The approach discussed in this subsection
also allows to incorporate scalar potentials
(see reference~\cite{kokado} for applications to the Hall effect).

We note
that substitution of the  symmetric gauge $A_i =  -{1 \over 2} \, \Bc
\varepsilon_{ij} x_j$ into $\check \psi$,
as given by equation (\ref{mattf}),
leads to
\[
\check \psi (\psi, A; \theta)
= \psi + {e \theta \Bc \over 4} \, \vec x \cdot \vec{\nabla}  \, \psi
+ {\cal O} (\theta ^2 )
\, .
\]
Here, the first order term in $\theta$ amounts to a $\theta$-dependent
scale transformation of $\psi$, which shows that the Seiberg-Witten
map (\ref{mattf}) is not unitary~\cite{scholtz}.
This map
may eventually be unitarized, see references~\cite{kokado,scholtz}.
Different aspects of the Seiberg-Witten map (like singularities or
ambiguities in the parametrization)
are discussed in the
work~\cite{suo}.

{\em In summary:}
The approach of Seiberg-Witten to NCQM amounts to considering
star products, and to assuming that
the fields of  NCQM
(i.e. $\check \psi, \check A_i, \check \lambda, \dots$)
and the coupling constants
of  NCQM  (i.e. $e_{{\rm SW}}$ and $ m_{{\rm SW}}$) are specific functions of
those occurring in ordinary quantum mechanics.
The star product is tantamount to  the symmetric gauge
for $\hat{\vec X}$ and thereby this approach
to a constant magnetic field
only works in a simple way
if the symmetric gauge is also chosen for $\check{\vec A}$.

\section{Summary and concluding remarks}

\noindent
{\bf Magnetic fields in NCQM -- Summary:}
For ${[ \hx_1 , \hx_2 ]} = \ri \theta \Id$,
the coupling to a {\bf constant magnetic field} can either be described
by considering a free Hamiltonian
$H = {1 \over 2m} \, \hat{\vec P} ^2 $
and a non-trivial commutator
${[ \hp_1 , \hp_2 ]} = \ri e B \Id$, or by considering
a trivial commutator ${[ \hp_1 , \hp_2 ]} =0$
and a Hamiltonian involving a vector potential.
For the latter case, we presented three approaches
(subsections~\ref{ahat}, \ref{gaugepot} and \ref{swmap})
all of which lead to a noncommutative field strength.
The vector potential or field strength appearing in these  approaches
can be expressed in various ways
in terms of the constant $\theta$-independent
field $B$.
By choosing this dependence on $B$
in an appropriate way, one recovers the spectral results of the
former formalism
(where ${[ \hp_1 , \hp_2 ]} = \ri e B \Id$).
The issue of gauge invariance of the energy spectrum
is  unproblematic in the operatorial
formulation (section~\ref{ahat})
and somewhat subtle in the other approaches.

The three approaches based on vector potentials 
lead to a non-canonical commutator 
between the coordinates and the kinematical momentum 
-- see equation (\ref{ncgal}). 
Thereby, they 
allow
for a treatment of a {\bf non-constant magnetic
field} to be identified with the
noncommutative field strength $F_{12}$.
However, the physical interpretation is
somewhat subtle due to the fact
that the field $F_{12}$ is not gauge invariant in this case.

\smallskip

\noindent
{\bf Concluding remarks:}
In these notes, we discussed some general properties 
of NCQM while limiting ourselves to two dimensions 
and to the case of a canonical commutator between 
$\hx_i$ and $\hp_j$: 
$[ \hx_i, \hp_j] = \ri \delta_{ij} \Id$.
The case of higher dimensions and of a more general 
algebra are to be addressed elsewhere~\cite{inprepa} 
along with other physical systems involving both 
scalar and vector potentials.
In that context, we will also review other methods of quantization
like the path integral approach to NCQM.


\vskip 1.2truecm


{\bf \Large Acknowledgments}

\vspace{3mm}


It is a great pleasure to thank
H.~Grosse,
P.~Exner, M.~Kibler and S.~Richard for valuable discussions.
F.G. thanks M.~Rausch de Traubenberg, M.~Sampaio,
J.~Dittrich and V.~Rivasseau
for the kind invitations to deliver talks
on the present subject
at the conferences
``Rencontres Alsaciennes
Math-Physique'' (Strasbourg),
``International Meeting on Topics in Quantum Field Theory''
(Belo Horizonte),
 ``Operator Theory in Quantum Physics'' (Prague)
and ``Non-commutative Geometry and Physics'' (Orsay).
 We also acknowledge the discussions on noncommutative field theories
with A.~Deandrea, C.~Ga\c{w}edzki, J.~Grimstrup, N.~Mamoudi and P.~Schuck.

\newpage

\newpage
\appendix
\section{Noncommuting coordinates in the truncated Landau problem
(Peierls' substitution)}

In the following, we will show that noncommuting coordinates
naturally appear in the Landau problem
in the limit of a {\em very strong magnetic field}~\cite{peierls, dunne}.
Since the separation of Landau levels (\ref{speclandau})
is of order $B/m$,
a large magnetic field $B$ is equivalent to a small mass $m$
and the limit $m\to 0$ amounts to a restriction to
the lowest Landau level (LLL).
Before presenting this limiting procedure in a rigorous manner,
we provide a simple heuristic argument~\cite{dunne,jackiw}.
The classical dynamics of the Landau problem
(supplemented by a {\em weak} scalar potential $V(x_1, x_2)$
describing impurities in the plane) is described
by the Lagrangian
\begin{eqnarray}
\label{6}
L_m \!\!\! & = &  \!\!\! \ds{1 \over 2} \, m \dot{\vec x} ^{\, 2}
+ \ds{e \over c} \, \vec A \cdot \dot{\vec x}  - V ({\vec x} \, )
\\
\!\!\! & = &  \!\!\!
 \ds{1 \over 2} \, m (\dot{x} _1 ^{ 2} +  \dot{x} _2 ^{ 2})
+ \ds{eB \over c} \, x_1  \dot{x} _2 - V ( x_1, x_2)
\, ,
\nonumber
\end{eqnarray}
where we have chosen the Landau gauge
$(A_1 , A_2) = (0 , Bx_1)$.
In the limit $m\to 0$, we are left with
\[
L_0 = \ds{eB \over c} \, x_1  \dot{x} _2 - V ( x_1, x_2)
\, ,
\]
i.e. an expression of the form
$p \dot{q} - H_0 (p,q)$
with a pair of canonically conjugate variables
$(p,q) = ({eB \over c} x_1, x_2)$
and a Hamiltonian $H_0 = V$.
Upon quantization, the classical variables become
operators $ {eB \over c} \hx_1$ and $\hx_2$ satisfying
CCR's,
$[ {eB \over c} \hx_1 , \hx_2]= -\ri \hbar \, \Id$, i.e.
\begin{equation}
\label{pei}
[ \hx_1 , \hx_2]= -\ri \, \ds{\hbar c \over eB} \, \Id
\, .
\end{equation}
Thus, {\em the restriction of the particle dynamics
to the LLL amounts to introducing noncommuting coordinates
in the plane} which satisfy the algebra (\ref{pei}),
the reduced Hamiltonian
$H_0 = V(\hx_1, \hx_2)$ describing this dynamics
being given by the potential
depending on the noncommuting  coordinates.
This so-called {\bf Peierls' substitution}
means that the effect of an impurity
(as described by the potential $V$  in the Lagrangian (\ref{6}))
on the eigenstate of the  LLL
can be evaluated to lowest order
by computing the eigenvalues $\epsilon_n$
of the effective Hamiltonian
$H_0 = V(\hx_1, \hx_2)$, where the commutator of $\hx_1$ and $\hx_2$
is proportional to $1/B$. More explicitly, if
$V(\hx_1, \hx_2) |n\rangle = \epsilon_n |n\rangle$,
then the energy levels for the quantum system associated to
(\ref{6}) are given by
$E_n = {1 \over 2} \, \hbar \omega_B + \epsilon_n$
(with $\omega_B \equiv \ds{|eB| \over mc}$)
in the approximation of strong $B$ and weak $V$
\cite{peierls, dunne, jackiw}.
(Incidentally, the result (\ref{pei}) coincides
-- up to a sign -- with the commutator $[\hx_1, \hx_2]$
that appears in the algebra (\ref{thb})
at the singular point
$\kappa \equiv 1 - {e \over \hbar c} \, B \theta = 0$.)

To make the previous argument more precise~\cite{andre}
(see also~\cite{jackiw,ouvry, magro}),
we consider the Hamiltonian operator
$H= {1 \over 2m} \sum_{j=1}^2 (P_j - {e \over c} A_j)^2$
as written in the symmetric gauge.
Recall that the eigenstate $|n,k \rangle$ of $H$
associated to the $n$-th Landau level
$E_n = \hbar \omega_B (n+ {1 \over 2})$
is infinitely
degenerate: $k \in \br$.
The projection operator
${\cal P}_n = \int_{\br} dk \, |n,k \rangle \, \langle n,k|$
onto the corresponding eigenspace  
can be expressed as follows in terms of the Hamiltonian
$H = \sum_{n=0}^{\infty} E_n {\cal P} _n$ \cite{andre}:
\[
{\cal P} _n = \ds{4 \over \pi (2n+1)} \,
\ds{
\sin
\left[
(n+ {1 \over 2} ) \, \pi \,
\left( H_n - \Id \right)
\right]
\over
\left( H_n - \Id \right)
\left( H_n + \Id \right)
}
\, , \qquad {\rm with} \ \;
H_n
\equiv
{1 \over E_n} \, H 
\, .
\]
Rather than truncating the theory to the LLL,
we may consider more generally the theory obtained by
{\em cutting off at an energy}
$E$ with $E_N \leq E < E_{N+1}$,
where $N\in \{ 0,1,2, \dots \}$ is fixed~\cite{ouvry,magro,andre}.
The operator
$\Pi_N = \sum_{n=0}^{N} {\cal P} _n$ then projects
onto the corresponding subspace of Hilbert space $\hs$.
In particular, for any operator $B$ on $\hs$, the expression
\[
\hat B \equiv \Pi_N B \Pi_N
\]
represents the truncation of $B$.
The authors of reference~\cite{andre}
determined the {\em commutators
of the truncated canonical operators}:
\begin{eqnarray}
\label{trunccom}
[ \hx_1 , \hx_2 ]
 \!\!\! & = &  \!\!\!
 -\ri \, \ds{\hbar c \over eB} \, (N+1) \, {\cal P} _N
\\
{[ \hp _1 , \hp _2 ]}
 \!\!\! & = &  \!\!\!
 -\ri \hbar \, \ds{eB \over 4 c} \, (N+1) \, {\cal P} _N
\nonumber
\\
{ [ \hx _i , \hp _j ] }
 \!\!\! & = &  \!\!\!
\ri \hbar \,  \delta_{ij} \, [ 1 - {1 \over 2}  (N+1)] \, {\cal P}_N
\, .
\nonumber
\end{eqnarray}
In particular, the first equation implies that the
LLL matrix elements of $[ \hx_1 , \hx_2]$ are given by
\[
\langle 0,k| \, [ \hx_1 , \hx_2] \, | 0, k^{\prime} \rangle =
 -\ri \, \ds{\hbar c \over e B}
\, \langle 0,k| 0, k^{\prime} \rangle
\,
\]
in agreement with the result (\ref{pei})
suggested by an heuristic argument.
Thus, the set of equations (\ref{trunccom}) gives a precise mathematical
meaning to the procedure of `projection to LLL'.
For $N \to \infty$, one recovers the canonical commutators~\cite{magro}.

The modifications to (\ref{trunccom})
brought about an interaction potential
$V_{\lambda} ( x_1^2 + x_2 ^2)$ are discussed in reference~\cite{scholtzcond}.
In general, the commutator $[ \hx_1 , \hx_2 ]$ then depends
on the parameter $\lambda$ characterizing the potential
$V_{\lambda}$.


\end{document}